\documentclass[11pt]{iopart}
\usepackage{bm}
\usepackage{hyperref}
\usepackage{amssymb}
\usepackage{feynmp}
\usepackage{amsopn}
\usepackage{setstack}
\RequirePackage{ifpdf}

\ifpdf
\else 
\fi

\newcommand{\Ps}{\mathcal{P}}
\newcommand{\Mp}{M_{\mathrm{P}}}

\newcommand{\E}{\mathbb{E}}
\newcommand{\ktrans}{k_{\perp}}
\newcommand{\vectktrans}{\vect{k}_{\perp}}
\newcommand{\Mpi}{M_{\pi}}
\newcommand{\LambdaQCD}{\Lambda_{\mathrm{QCD}}}
\newcommand{\frag}[2]{D_{{#1}}^{\mathrm{{#2}}}}
\newcommand{\nplus}{N^{+}}
\newcommand{\nminus}{N^{-}}

\renewcommand{\d}{\mathrm{d}}
\newcommand{\vect}[1]{\bm{\mathrm{{#1}}}}
\renewcommand{\e}[1]{\mathrm{e}^{{#1}}}
\newcommand{\im}{\mathrm{i}}

\newcommand{\borel}{\mathcal{B}}

\renewcommand{\leq}{\leqslant}
\renewcommand{\geq}{\geqslant}

\makeatletter
\newcommand\numberwithin[2]{\@addtoreset{#1}{#2}}
\makeatother
\numberwithin{footnote}{section}

\begin{document}
	\title{A parton picture of de Sitter space during slow-roll inflation}
	\date{\today}
	\author{David Seery}
	\address{Department of Applied Mathematics and Theoretical Physics \\
	Wilberforce Road, Cambridge, CB3 0WA, United Kingdom}
	\eads{\mailto{djs61@cam.ac.uk}}
	\submitto{JCAP}
	\begin{abstract}
	It is well-known that expectation values in de Sitter space are
	afflicted by infra-red divergences. Long ago, Starobinsky proposed that
	infra-red effects in de Sitter space could be accommodated by
	evolving the long-wavelength part of the field according to the
	classical equations of motion plus a stochastic source term.
	I argue that---%
	when quantum-mechanical loop corrections are taken into account---%
	the separate universe picture of superhorizon evolution in de
	Sitter space is equivalent, in a certain
	leading-logarithm approximation, to Starobinsky's stochastic approach.
	In particular the time evolution of a box of de Sitter space
	can be understood in exact analogy with the DGLAP evolution of
	partons within a hadron, which
	describes a slow logarithmic evolution
	in the distribution of the hadron's constituent partons
	with the energy scale at which they are probed.

	\vspace{3mm}
	\begin{flushleft}
	\end{flushleft}
	\end{abstract}
	\maketitle
	
	\begin{fmffile}{diags}

	\section{Introduction}

	In the years since the WMAP satellite provided the first precise
	measurement of the angular power spectrum of the
	cosmic microwave background (CMB), a
	consensus has emerged in which
	inflation is
	the most likely candidate for the origin
	of the primordial density perturbation. If this is true it will
	almost certainly require the presence of new degrees of freedom
	at a scale associated with the inflationary era,
	perhaps in the region $10^{12}$~--~$10^{16}$ GeV,
	and would represent a remarkable discovery of new physics at an energy
	at least $10^8$ times the accessible limit at the
	Large Hadron Collider and 
	far beyond the reach of any terrestrial experiment.
	
	The prospect of discoveries such as these
	has led to significant investment in microwave
	background experiments, which are now approaching the sensitivity
	required to detect effects taking place at subleading order in
	perturbation theory \cite{Yadav:2007yy,Komatsu:2008hk}.
	These effects are expected to play a crucial
	role in discriminating among the competing theories which could account
	for the gross Gaussian, adiabatic and scale-invariant character
	of the inflationary density perturbation.
	Indeed, one immediate consequence
	of our imminent ability to measure such tiny contributions
	has been a strong pressure to develop and refine the
	theory of inflationary perturbations, which underlies the interpretation
	of all CMB observations. At a practical level this has led to the
	availability of full predictions for the non-linearity in the
	three- and four-point correlation functions of the primordial
	curvature perturbation in single-field inflation
	\cite{Maldacena:2002vr,Lyth:2005fi,Seery:2006vu,Seery:2008ax}
	and partial predictions in multi-field models
	\cite{Seery:2005gb,Vernizzi:2006ve,Kim:2006te,Battefeld:2006sz,
	Seery:2006js,Byrnes:2006vq}.
	At a more conceptual level, interesting new approaches have been
	developed by borrowing the idea of an \emph{effective field
	theory} from particle physics
	\cite{Cheung:2007st,LeBlond:2008gg,Weinberg:2008hq,ArmendarizPicon:2008yv,
	Weinberg:2008mc,Weinberg:2008nf,Weinberg:2008si,Shandera:2008ai}.
	These enable us to ask fundamental questions about theories of
	inflation without making a commitment to any specific model.
	
	An example of such a fundamental question concerns the validity
	of perturbation theory. During inflation there are at least two
	interesting perturbative scales. The first is a measure of the
	deceleration of the Hubble rate, defined by
	$\epsilon \equiv - \dot{H}/H^2$, which provides an indication of
	the time scale over which the vacuum expectation values of background
	fields are coherently evolving due to macroscopic classical effects.
	The second is the ratio of the Hubble scale to the Planck scale,
	$H/\Mp$, which determines the importance of higher orders in
	perturbation theory associated with ``quantum'' corrections.
	Our ability to extract predictions from any theory of inflation
	is usually limited to the lowest few orders in both $\epsilon$
	and $H/\Mp$, or other related small quantities. Unfortunately, it has
	been known for a long time that such predictions can be afflicted
	with infra-red divergences which compensate for the smallness
	of these expansion parameters and spoil our ability to perform
	meaningful calculations
	\cite{Sasaki:1992ux,Suzuki:1992gi}.
	These divergences have been explored in a series of papers
	by Woodard and collaborators: see, for example,
	Refs.~\cite{Tsamis:2005hd,Miao:2006pn,Prokopec:2007ak,
	Tsamis:2008it} which contain references to the earlier literature.
	
	We would like to be sure that when we make predictions
	which are to be compared with precision CMB data, we obtain the
	right answer for the right reason. To achieve this confidence
	in our quantitative predictions,
	and the methods used to obtain them,
	it is important to arrive at a clear understanding of
	infra-red issues. Accordingly, they have been subject to
	investigation by many authors
	\cite{Mukhanov:1996ak,Unruh:1998ic,
	Boubekeur:2005fj,Weinberg:2005vy,Losic:2005vg,
	Zaballa:2006pv,
	Sloth:2006az,Losic:2006ht,Weinberg:2006ac,Sloth:2006nu,
	Lyth:2007jh,Seery:2007we,Seery:2007wf,Losic:2007tu,
	Urakawa:2008rb,Unruh:2008zz,Losic:2008ht,
	Dimastrogiovanni:2008af,Adshead:2008gk,Urakawa:2009my}.
	The presence of large infra-red effects is symptomatic of
	computing an observable defined on some particular length scale
	within a much larger patch of de Sitter space
	\cite{Lyth:2007jh,Seery:2007wf,Bartolo:2007ti,Enqvist:2008kt},
	leading to the existence of a large hierarchy of scales.
	It is the logarithm of this hierarchy which enters in conjunction
	with the scales $\epsilon$ and $H/\Mp$, and spoils
	na\"{\i}ve perturbation theory.
	Whenever large logarithms of this sort
	play a significant role in quantum field theory, it is usually
	the case that their resummation can be
	described by the renormalization group equation.
	We should therefore expect that significant infra-red effects, if they
	exist, can be accommodated within this framework
	\cite{Boubekeur:2005fj,Weinberg:2005vy,
	Lyth:2006gd,Lyth:2007jh,Byrnes:2007tm,Seery:2007wf,Podolsky:2008qq}.
	On the other hand, it has been suggested by many
	authors that the stochastic
	approach to inflation, originally pioneered by
	Starobinsky \cite{Starobinsky:1986fx,Starobinsky:1994bd},
	functions as an infra-red regulator and leads to
	infra-red finite predictions
	\cite{Tsamis:2005hd,Miao:2006pn,
	Bartolo:2007ti,Enqvist:2008kt,Podolsky:2008du,Riotto:2008mv}.
	In this interpretation, the presence of large infra-red terms should
	be understood to reflect the potential for large fluctuations
	to build up between widely separated points in de Sitter space,
	and eventually a sensitivity to the onset of eternal inflation.
	This prescription was suggested earlier in
	Refs.~\cite{Tsamis:2005hd,Miao:2006pn}.
	
	This resolution of the problem of infra-red divergences, if correct,
	would be quite remarkable.
	However, it presents a number of puzzles.
	Firstly, we are used to applications in which
	the renormalization group equation leads to screening of
	masses and couplings constants in the ultra-violet,
	rather than some form of stochastic dynamics.
	It is not so easy to see how
	the two approaches could be related.
	Secondly, although one can check, once a stochastic formulation
	is available, that it correctly reproduces correct infra-red behaviour,
	it would be nice to have an argument which begins with
	the existence of large infra-red logarithms and arrives
	at the Langevin equation which describes Starobinsky's stochastic
	dynamics. For this purpose, one can seek analogies in other
	examples of field theory where infra-red effects play an important
	role. The key example of this type occurs in hadron physics,
	where the fundamental degrees of freedom belonging to the
	gauge theory of colour---%
	the quarks and gluons---are confined by soft QCD effects.
	
	In this paper I would like to suggest that
	although a large hierarchy of scales in de Sitter space can certainly
	be understood in terms of renormalization group flow, an even
	better analogy might be with the
	Dokshitzer--Gribov--Lipatov--Altarelli--Parisi (DGLAP) equation
	\cite{Dokshitzer:1977sg,
	Gribov:1972ri,Gribov:1972rt,Altarelli:1977zs,Kogut:1974ni}
	which describes the evolution of parton distribution functions
	within a hadron. These distribution functions satisfy an equation---%
	itself a manifestation of the renormalization group---%
	which endows them with a slow logarithmic variation as
	one changes the energy
	scale at which the target hadron is to be probed.
	In this analogy, one can show that the DGLAP equation
	\emph{precisely} reproduces
	the Fokker--Planck equation obtained by Starobinsky.
	This equation describes the
	diffusion of probability in time as a light scalar field in
	de Sitter space evolves in the presence of quantum effects.
	This Fokker--Planck equation could therefore be interpreted
	as \emph{the} de Sitter renormalization group equation.
	
	The structure of this paper is as follows.
	In \S\ref{sec:hadrons} the parton picture of hadron structure is
	briefly recalled, in a form which will allow easy generalization
	to de Sitter space. This generalization is performed in
	\S\S\ref{sec:desitter}--\ref{sec:predictions}.
	Finally, I conclude with a discussion in \S\ref{sec:discussion}.
	Two appendices supplement the discussion of the
	stochastic formulation of inflation
	which occurs in the main text. Starobinsky's original
	argument, based on a recursive coarse-graining of the
	Heisenberg field and the introduction of a Langevin equation,
	is recalled in \ref{sec:starobinsky}. In addition,
	to aid comparison of the argument given
	in the present paper with that of other authors,
	\ref{sec:fokkerplanck} gives two derivations of the
	corresponding Fokker--Planck equation, one based on
	It\={o}'s stochastic calculus and the other on a conventional
	path integral.
	Some issues tangential to the main discussion are briefly
	summarized in \ref{sec:subtleties}.
	
	Throughout this paper, units are chosen so that $\hbar = c = 1$
	and the reduced Planck mass (defined by
	$\Mp^{-2} \equiv 8 \pi G$, where $G$ is Newton's gravitational
	constant) is set to unity. The metric is chosen with sign
	convention $(-,+,+,+)$.
	
	\section{Hadrons and the parton picture}
	\label{sec:hadrons}
	
	\subsection{Deep inelastic scattering and parton
	distribution functions}
	
\begin{figure}
		\begin{center}
			\vspace{5mm}
			\begin{fmfgraph*}(150,100)
				\fmfleft{l1,l2}
				\fmfright{r}
				\fmftop{t}
				\fmfbottom{b}
				\fmfpen{thin}
				\fmf{electron,label=$k$,label.side=left}{l1,v1}
				\fmf{electron,label=$k'$}{v1,t}
				\fmf{photon,label=$Q$}{v1,v2}
				\fmf{quark,label=$p+q$}{v2,r}
				\fmf{quark}{l2,v3}
				\fmf{quark}{v3,b}
				\fmf{quark,label=$p$}{v3,v2}
				\fmfforce{(0,0.8h)}{l1}
				\fmfforce{(0,0.2h)}{l2}
				\fmfforce{(w,0.5h)}{r}
				\fmfforce{(0.7w,0)}{b}
				\fmfforce{(0.7w,h)}{t}
				\fmfforce{(0.5w,0.8h)}{v1}
				\fmfforce{(0.5w,0.2h)}{v3}
				\fmffreeze
				\fmfi{plain}{vpath (__l2,__v3) shifted (thick*(0,2))}
				\fmfi{plain,label=$P$,label.side=right}
				     {vpath (__l2,__v3) shifted (thick*(0,-2))}
				\fmfi{plain}{vpath (__v3,__b)  shifted (thick*(1,2))}
				\fmfi{plain}{vpath (__v3,__b)  shifted (thick*(-1,-2))}
				\fmfv{decoration.shape=circle,decoration.filled=shaded,
				      decoration.size=0.2h}{v3}
				\fmfv{label=$\mbox{\small electron}$,label.angle=180}{l1}
				\fmfv{label=$\mbox{\small hadron}$,label.angle=180}{l2}
				\fmfv{label=$\mbox{\small $\rightarrow$ hadron jet}$,
				      label.angle=0}{r}
			\end{fmfgraph*}
		\end{center}
		\caption{\label{fig:dis}Kinematics of deep inelastic
		electron--hadron scattering. A high energy electron impinges
		on the target hadron and interacts electromagnetically with
		one of its constituent partons. After the interaction, the
		original hadron is disrupted and the ejected quark materializes
		as a jet of hadrons collinear with the motion of the
		initial electron.}
	\end{figure}
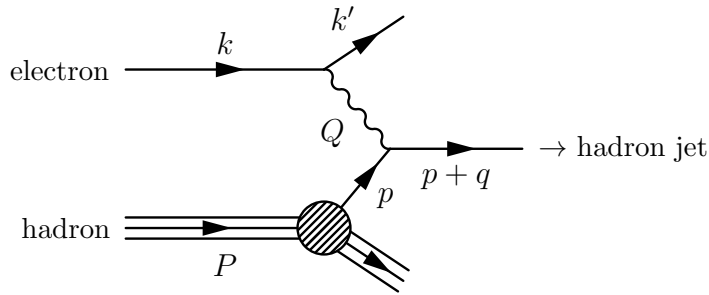

	The parton picture of hadron structure grew out of the analysis
	of so-called \emph{deep inelastic scattering} experiments
	(see Fig.~\ref{fig:dis}),
	which involved collisions between electrons and protons
	in which a large invariant momentum $Q^2$ was transferred from the
	$e^-$ beam to the target proton. In these experiments it was
	observed that the scattering rate behaved as if the electrons
	were interacting with an elementary electromagnetically charged
	fermion, whereas such hard scattering behaviour was virtually
	absent in direct proton--proton collisions. To explain these
	observations, Bjorken and Paschos \cite{Bjorken:1969ja}
	and Feynman \cite{Feynman:1973xc} proposed that hadrons could be
	understood as a loosely bound collection of
	pointlike constituents, called \emph{partons}. In this picture,
	the electron (which does not feel the strong force)
	does not interact with the hadron as a whole but rather scatters
	incoherently from one of its constituents.
	
	In modern terms the partons can be identified with quarks and
	gluons, which are the fundamental degrees of freedom of QCD.
	At low energies these are confined into colour-neutral states by soft
	processes, giving hadrons a characteristic size of order
	$\Mpi^{-1}$ where $\Mpi \sim \LambdaQCD$ is the pion mass and
	$\LambdaQCD \approx 200 \, \mbox{MeV}$ is the QCD scale.
	At energies much greater than $\LambdaQCD$ the quarks and gluons
	behave roughly like a plasma of free particles.
	
	The parton model can only be expected to supply a good
	approximation if the impinging electron is
	sufficiently energetic to resolve the internal structure of the
	target hadron. This internal structure is described by so-called
	parton distribution functions, labelled $f_i$ (but varying
	from hadron to hadron) for each species $i$
	which can be present, and
	defined so that $f_i(x) \, \d x$ is the probability of finding
	a constituent parton of species $i$ carrying a fraction
	$x$ ($0 \leq x \leq 1$) of the parent hadron's total momentum.
	These can be thought of as coarse-grainings over the hadron
	wavefunction.
	Since the partons are supposed to be bound within the hadron,
	their momenta transverse to its direction of propagation must all
	be small and to a good approximation
	the partons can be taken to move collinearly.
	The leading corrections to this picture will be suppressed by powers
	of $\ktrans / P$, where $\ktrans \equiv | \vectktrans |$ is of order
	the typical transverse 3-momentum (of order $\sim \Mpi$)
	and $P = | \vect{P} |$ is the 3-momentum of the parent hadron.
	
	Once the parton distribution functions are at our disposal,
	and taking into account the assumption that a probe scatters
	incoherently off a single constituent parton,
	it is clear how we can write the cross-section for a deep inelastic
	scattering event. In the case described in Fig.~\eref{fig:dis},
	where an electron of momentum $k$
	disrupts a hadron of momentum $P$, it follows that
	\begin{eqnarray}
		\fl\nonumber
		\sigma\Big\{ e^-(k) + p(P) \rightarrow e^-(k') + Y \Big\}
		\\ \mbox{}
		=
		\int_0^1 \d x
		\sum_j f_j(x) \sigma \Big\{
			e^-(k) + q_j(x P) \rightarrow e^-(k') + q_j(p')
		\Big\} ,
		\label{eq:parton-cross-section}
	\end{eqnarray}
	for any hadronic final state $Y$, where the sum over $j$ includes all
	quark flavours, with quarks and antiquarks contributing separately.
	Similar formulae can be obtained for any desired partonic interaction:
	the general scheme is always parallel to
	Eq.~\eref{eq:parton-cross-section}, which factorizes into
	a hard subprocess describing interactions among the partons
	and a soft parton distribution function. The hard subprocess is
	independent of the hadron in which the partons are contained,
	and the parton distribution function is independent of the
	interaction which takes place. In particular, the distributions
	$f_i$ depend only on $x$
	and are independent of the momentum transfer $Q^2$ which
	is carried by the intermediate boson in Fig.~\ref{fig:dis}.
	This behaviour is known as \emph{Bjorken scaling}
	\cite{Bjorken:1968dy}.
	
	A cross-section such as Eq.~\eref{eq:parton-cross-section} is called
	``inclusive,'' because it does not distinguish between the
	various allowed final hadronic states $Y$.
	It is sometimes possible to measure more ``exclusive''
	(or ``semi-inclusive'') rates which
	discriminate among the hadrons which can be present in $Y$.
	To describe such exclusive rates one can introduce
	fragmentation functions $\frag{i}{h}(z)$, which are the
	final-state analogues of the parton distribution functions $f_i(x)$:
	these give the probability for a parton of species $i$ to
	produce a hadron $\mathrm{h}$ in the final state which carries a
	fraction $z$ of the parent parton's momentum.
	An exclusive cross-section can be written in a form
	analogous to Eq.~\eref{eq:parton-cross-section}, using the
	fragmentation functions to sum the final-state products of the hard
	subprocess into final-state products of the overall hadronic
	process.
	
	The parton picture was introduced as a phenomenological model,
	with some basis in an intuitive understanding of hadron physics.
	As such it is independent of QCD itself, or more generally
	the existence of an underlying gauge theory which exhibits confinement
	in the infra-red.
	It is only
	the identification of partons with the fundamental excitations of
	an asymptotically free non-Abelian field theory which invests the model
	with real meaning. Indeed, the parton model can be formally
	derived from QCD \cite{Collins:1987pm}.
	
	\subsection{Parton evolution and the DGLAP equation}
	
	In the simplest parton model, Bjorken scaling is exact.
	However, if QCD
	processes are taken into account this is no longer true;
	instead, a tower of infra-red divergences appears which can
	compensate for the smallness of the QCD coupling constant at high
	energy. When resummed, these divergences predict slow violations
	of Bjorken scaling and give rise to evolution equations which
	control how the parton distributions functions $f_i$
	(and fragmentation functions $\frag{i}{h}$) evolve
	with the momentum scale $Q^2$ at which the hadron is to be probed.
	
	What is the origin of these QCD effects? Consider any process with
	outgoing quarks. At the point where these leave the diagram,
	any such process must contain a vertex of the form
	\begin{center}
		\parbox{15mm}
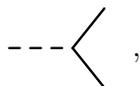
{
		\begin{fmfgraph*}(40,30)
			\fmfleft{l}
			\fmfright{r1,r2}
			\fmf{dashes}{l,v}
			\fmf{plain}{v,r1}
			\fmf{plain}{v,r2}
		\end{fmfgraph*}} ,
	\end{center}
	where the dashed line is connected to the rest of the diagram.
	However, nothing can prevent the outgoing quarks from radiating
	gluons, which means that at leading order in
	the strong coupling constant, $\alpha_s$, we must also
	include diagrams of the form
	\begin{center}
		\parbox{15mm}
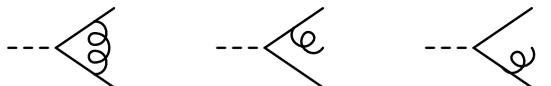
{
		\begin{fmfgraph*}(40,30)
			\fmfleft{l}
			\fmfright{r1,r2}
			\fmf{dashes}{l,v}
			\fmf{plain,tension=0.5}{v,v1}
			\fmf{plain}{v1,r1}
			\fmf{plain,tension=0.5}{v,v2}
			\fmf{plain}{v2,r2}
			\fmfforce{(0,0.5h)}{l}
			\fmfforce{(w,h)}{r1}
			\fmfforce{(w,0)}{r2}
			\fmfforce{(0.45w,0.5h)}{v}
			\fmffreeze
			\fmf{gluon}{v2,v1}
		\end{fmfgraph*}}
		\hspace{1cm}
		\parbox{15mm}{
		\begin{fmfgraph*}(40,30)
			\fmfleft{l}
			\fmfright{r1,r2,g}
			\fmf{dashes}{l,v}
			\fmf{plain}{v,v1,r1}
			\fmf{plain}{v,r2}
			\fmfforce{(0,0.5h)}{l}
			\fmfforce{(w,h)}{r1}
			\fmfforce{(w,0)}{r2}
			\fmfforce{(0.45w,0.5h)}{v}
			\fmfforce{(w,0.5h)}{g}
			\fmffreeze
			\fmf{gluon}{v1,g}
		\end{fmfgraph*}}
		\hspace{1cm}
		\parbox{15mm}{
		\begin{fmfgraph*}(40,30)
			\fmfleft{l}
			\fmfright{r1,r2,g}			
			\fmf{dashes}{l,v}
			\fmf{plain}{v,r1}
			\fmf{plain}{v,v1,r2}
			\fmfforce{(0,0.5h)}{l}
			\fmfforce{(w,h)}{r1}
			\fmfforce{(w,0)}{r2}
			\fmfforce{(0.45w,0.5h)}{v}
			\fmfforce{(w,0.5h)}{g}
			\fmffreeze
			\fmf{gluon}{v1,g}
		\end{fmfgraph*}} .
	\end{center}
	The first of these diagrams is a loop correction, which can produce
	a divergence when the momentum carried by the circulating gluon
	approaches zero. The remaining two diagrams account for gluon
	radiation from the outgoing quarks, and include soft gluons in
	the final state. If $\alpha_s$ is small, these diagrams will be
	suppressed compared to the undressed diagram without gluon radiation.
	Unfortunately,
	however, these gluon-corrected diagrams may be singular when the
	momenta carried by the final-state gluons approaches zero, or where they
	are emitted collinearly with the outgoing quark. In some cases
	the divergences produced by the combination of these diagrams
	cancel, but there is no reason of principle for this to occur.
	Similar corrections must be taken into account for ingoing particles.
	Where cancellation does not
	occur, any left-over divergences may
	overcome the smallness of $\alpha_s$ and
	cause the parton content encountered by an impinging hard
	particle to evolve with $Q^2$.
	
	The physical meaning of this evolution is simple to understand.
	An isolated quark propagating according to the usual rules of
	quantum field theory is not alone, but rather is accompanied by
	a cloud of virtual particles which are constantly being emitted
	and re-absorbed by the physical quark. Any projectile which
	strikes the cloud with sufficiently high energy has an opportunity to
	resolve and interact with one of the virtual particles in its interior,
	rather than the parent quark. As $Q^2$ increases, such a projectile
	can resolve vacuum fluctuations with increasingly short lifetimes.
	It follows that the composition of the virtual cloud must exhibit
	a slow variation with $Q^2$.
	Indeed, we can imagine it to be governed
	by a system of equations of Boltzmann type, which describe
	a sort of equilibrium
	among the various species of particle which can exist within the cloud.
	These are the DGLAP equations, sometimes known
	(especially in the older literature) as the
	Altarelli--Parisi equations. In the context of QCD they were
	obtained by Altarelli \& Parisi \cite{Altarelli:1977zs} using
	the method of the operator product expansion.
	Their interpretation as an
	approximate Boltzmann system was given later by Collins \&
	Qiu \cite{Collins:1988wj}.
	
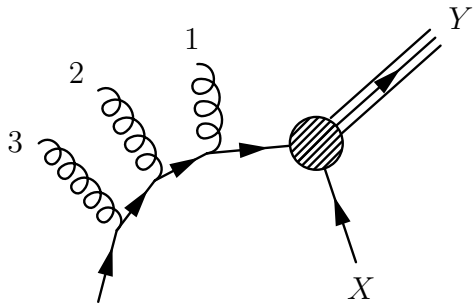
\begin{figure}
		\vspace{5mm}
		\begin{center}
			\begin{fmfgraph*}(150,100)
				\fmftop{t1,t2}
				\fmfleft{l}
				\fmfbottom{b1,b2}
				\fmfright{r}
				\fmf{quark,tension=4}{b1,v1}
				\fmf{quark,tension=4}{v1,v2}
				\fmf{quark,tension=4}{v2,v3}
				\fmf{quark,tension=2}{v3,v4}
				\fmf{gluon}{v1,l}
				\fmf{gluon}{v2,t1}
				\fmf{gluon}{v3,t2}
				\fmf{quark}{b2,v4}
				\fmf{quark}{v4,r}
				\fmfforce{(0.15w,0)}{b1}
				\fmfforce{(0,0.6h)}{l}
				\fmfforce{(0.15w,0.8h)}{t1}
				\fmfforce{(0.4w,0.9h)}{t2}
				\fmfforce{(0.8w,0.15h)}{b2}
				\fmfforce{(w,h)}{r}
				\fmfforce{(0.7w,0.6h)}{v4}
				\fmffreeze
				\fmfv{decoration.shape=circle,decoration.filled=shaded,
			      	decoration.size=0.2h}{v4}
				\fmfi{plain}{vpath (__v4,__r) shifted (thick*(-1,1.5))}
				\fmfi{plain}{vpath (__v4,__r) shifted (thick*(1,-1.5))}
				\fmfv{label=$X$,label.angle=-70}{b2}
				\fmfv{label=$Y$}{r}
				\fmfv{label=$3$}{l}
				\fmfv{label=$2$}{t1}
				\fmfv{label=$1$}{t2}
			\end{fmfgraph*}
			\caption{\label{fig:braking-radiation}A parton brakes as it
			enters into a collision with some incoming particle $X$,
			before
			scattering into a final hadronic state $Y$. As it brakes,
			it radiates an arbitrary number of
			soft gluons $\{ \cdots, 3, 2, 1 \}$ and moves increasingly
			off-shell. The impinging $X$ projectile can resolve this
			emission cascade if it is sufficiently energetic.}
		\end{center}
	\end{figure}

	Consider any parton which is destined to interact with some
	impinging projectile $X$. If $X$ is sufficiently energetic, it is
	possible to resolve processes by which the parton brakes into
	(or out of) the
	collision through emission---bremsstrahlung---of
	soft quanta, as described in
	Fig.~\ref{fig:braking-radiation}.
	If the parton is moving increasingly off-shell as it passes from
	the outside of the emission chain towards the collision region,
	then each radiation event can be accompanied by a large infra-red
	logarithm. It follows that to obtain a meaningful picture of this
	cascade of emission events, we must resum all diagrams with the form
	of Fig.~\ref{fig:braking-radiation} which describe
	radiation of an arbitrary number of soft quanta.
	
	Any diagram of this type, involving radiation of $N$ quanta, can be
	built by sewing together $N$ copies for the diagram for emission of
	a single quantum. To deal with this, one can write
	the probability for an incoming parton
	(say of species $i$) to resolve into an outgoing
	parton of species $j$ with a fraction $z$ of its original momentum,
	accounting for radiation of a single
	soft particle with transverse energy $\delta \ln Q^2$,
	in terms of the so-called
	\emph{Altarelli--Parisi splitting function},
	$P_{j\leftarrow i}(z)$, which satisfies
	\begin{equation}
		P_{j\leftarrow i}(z) = \delta(1-z) + \alpha_s p(z) \delta \ln Q^2
		\label{eq:splitting-function}
	\end{equation}
	for some $p(z)$ which can be computed by studying $S$-matrix
	elements according to the usual rules of quantum field theory.%
		\footnote{Note that $p(z)$ may not be a pure function;
		in order to arrive at a properly normalized $P_{ii}$, it may be
		necessary to interpret $p(z)$ as a distribution in its own right
		by adding some admixture of $\delta(1-z)$.}
	In writing Eq.~\eref{eq:splitting-function} and in what follows,
	it has been assumed that $\delta \ln Q^2 \ll 1$ and
	that terms of order $(\delta \ln Q^2)^2$ or smaller are negligible.
	One can understand the form of this equation
	by observing that if $\delta \ln Q^2 = 0$, so that no soft particle is
	radiated, then the original parton must remain at its initial momentum.
	This accounts for the leading term $\delta(1-z)$. On the other hand,
	if a particle of momentum $\delta \ln Q^2$ is radiated, it is possible
	for the original parton to downgrade its momentum to some
	fraction $z \neq 1$. The potential for this degradation is suppressed
	by the coupling $\alpha_s$, but can be enhanced if
	the available phase space $\delta \ln Q^2$ is large.
	
	To avoid complications,
	let us restrict attention to a single parton species $i$ which does
	not undergo transmutation to or from any other species.
	The parton distribution function $f_i(x)$
	evolves according to the master equation
	\begin{equation}
		f_i(x; Q^2 + \delta Q^2) =
		\int_0^1 \d x' \int_0^1 \d z \;
			P_{i\leftarrow i}(z) f_i(x'; Q^2) \delta(x - x' z) .
		\label{eq:altarelli-parisi}
	\end{equation}
	This is a principle of detailed balance, or
	Chapman--Kolmogorov equation, in which we account for all the ways a
	parton could arrive at momentum fraction $x$ at the probe scale
	$Q^2 + \delta Q^2$ by summing over an intermediate step at
	a probe scale $Q^2$.
	Passing to the continuum limit, one arrives at an integro-differential
	equation which describes the evolution of $f_i$ with $\ln Q^2$,
	\begin{equation}
		\frac{\partial f_i(x; Q^2)}{\partial( \alpha_s \ln Q^2 )}
		= \int_x^1 \frac{\d z}{z} \; p(z) f_i(x/z; Q^2) .
		\label{eq:DGLAP}
	\end{equation}
	This is the prototype DGLAP equation.
	It has the characteristic form of a renormalization group equation
	in $\ln Q^2$. If other species of parton, say of type $j$, can evolve into
	$i$-partons by soft emission then one must include splitting functions
	of the form $P_{j\leftarrow i}$ in
	Eqs.~\eref{eq:altarelli-parisi}--\eref{eq:DGLAP}
	which couple $f_i(x)$ to the relevant $f_j(x)$.
	By this process one arrives at a system of equations which can
	be thought of as an approximate Boltzmann hierarchy
	with collision integrals given by the right-hand sides
	of Eqs.~\eref{eq:altarelli-parisi}--\eref{eq:DGLAP}.
	Accounts of all these complexities can be found in the literature;
	see, for example, Refs.
	\cite{Dissertori:2003pj,Leader:1996hm,Peskin:1995ev}.
	
	Eq.~\eref{eq:DGLAP} must be supplemented with an appropriate boundary
	condition which determines the starting point for the DGLAP evolution.
	For partons confined within hadrons, such as quarks and gluons,
	a boundary condition cannot be obtained from first principles
	because the distribution functions $f_i$ depend on soft QCD effects
	which are presently incalculable. Instead, a set of distribution
	functions at some given scale must be extracted from experiment
	and
	the DGLAP evolution of these distribution functions can be compared
	with the observed distributions at a different scale.
	At present, this approach gives
	reasonable agreement between theory and observation. On the other hand,
	the foregoing discussion applies just as well for particles which
	can exist in isolation, such as an electron. For such particles,
	it is easy to find an appropriate boundary condition.
	For example, when probed at a scale corresponding to its own
	Compton wavelength, an electron should resolve only into itself
	and not any member of its surrounding cloud. Therefore
	we should set
	$f_e(x,Q^2) = \delta(1-x)$ at $Q^2 \sim m_e^2$, where $m_e$ is the
	electron mass.
	
	A similar discussion can be given for the fragmentation functions
	$\frag{i}{h}$. To leading order in $\alpha_s$ the splitting functions
	$P_{j \leftarrow i}$ are the same for both, although differences occur
	at higher orders in perturbation theory. In particular, the
	fragmentation functions also evolve according to a DGLAP equation
	which has the form of Eq.~\eref{eq:DGLAP}.
	
	\subsection{Leading logarithms}
	
	What has been achieved by using Eq.~\eref{eq:DGLAP} to evolve
	the parton distribution functions between two widely separated
	probe scales (say $Q^2$ and $Q_0^2$)? Although
	Eq.~\eref{eq:altarelli-parisi} gives a correct accounting
	of all powers of $\delta \ln Q^2$ terms in the continuum limit,
	it is not necessarily exact because the probability function $p(z)$
	must be computed by assembling Feynman diagrams into
	S-matrix elements. It is therefore a perturbative expansion in
	powers of the strong coupling constant, $\alpha_s$,
	and perhaps other small quantities.
	If we work to leading order in $\alpha_s$, the solution of the
	DGLAP equation will account correctly for all terms of
	the form $(\alpha_s \ln Q^2/Q_0^2)^n$. However, it will \emph{not} give
	useful information regarding
	terms which are suppressed by higher powers of
	$\alpha_s$, such as $\alpha_s^n \ln^{n-1} Q^2/Q_0^2$.
	This level of precision
	is known as the \emph{leading-logarithm approximation},
	sometimes abbreviated as ``LLA''. Better approximations,
	which account for terms suppressed by extra powers of $\alpha_s$,
	can be found by retaining higher powers of the coupling constant in
	$p(z)$.
	
	The leading-logarithm approximation has a particular interpretation
	in terms of the emission chain depicted in
	Fig.~\ref{fig:braking-radiation}. Terms which contribute at
	leading logarithmic order arise from the region of phase space
	where the cascade of emitted quanta is \emph{strongly ordered}
	\cite{Dissertori:2003pj,Leader:1996hm,Peskin:1995ev}, in the
	sense that
	\begin{equation}
		Q_0^2 \ll Q_n^2 \ll \cdots \ll Q_3^2 \ll Q_2^2 \ll Q_1^2 \ll Q^2 .
		\label{eq:strong-order}
	\end{equation}
	where $Q_n^2$ is the invariant transverse momentum carried away by
	the $n$th quantum radiated in the cascade, remembering that
	the label `1' is attached to the quantum closest to the collision
	region whereas quanta labelled by higher $n$ are increasingly distant.
	Other configurations of radiative cascade are possible,
	such as emission of two or more particles at roughly equal $Q^2$, but
	since there is only a single logarithm for each hierarchy
	in $Q^2$ and each emission costs a factor $\alpha_s$ these configurations
	contribute at the next-to-leading logarithm approximation
	(``NLLA'') or lower.
	
	The leading logarithm resummation only gives a meaningful approximation
	if all significant effects actually arise from the large logarithms
	under consideration. In hadron physics, large effects can also arise
	from evolution at small Bjorken-$x$, where $\ln x$ logarithms can make
	a dominant contribution to the physics. In this region,
	the DGLAP equation must be
	replaced by another renormalization group equation, the so-called
	JIMWLK equation \cite{Weigert:2005us}.
	I shall return to this problem in \S\ref{sec:discussion}.

	\section{A parton picture of de Sitter space}
	\label{sec:desitter}
	
	Now let us apply these ideas to a box of de Sitter space which is
	undergoing slow-roll inflation. Remarkably, it will be possible to
	find analogues for all the concepts which played a role in the
	discussion of hadrons and parton evolution outlined
	in \S\ref{sec:hadrons}, including the parton distribution
	and fragmentation functions, and the strongly-ordered radiative cascade.
	The most significant difference is that
	the parton distribution is no longer probed by an impinging
	projectile such as an electron, because in calculating the
	density perturbation generated during inflation there is no analogue
	of a scattering event. Nevertheless, the concept of a probe is
	still implicitly present in the guise of sampling the density
	fluctuation smoothed on Hubble-sized regions. Indeed,
	the wavenumber corresponding to the Hubble scale
	satisfies $k = aH$ and it will transpire that
	$k$ plays the role of the probe momentum
	$\sqrt{Q^2}$, whereas the expectation values of the scalar fields
	which characterize the de Sitter phase
	play the role of the Bjorken variable $x$.
	
	These ideas will be developed in this section,
	in the course of which
	it will be possible to transcribe
	Eqs.~\eref{eq:parton-cross-section}--\eref{eq:DGLAP}
	and the strong ordering condition
	\eref{eq:strong-order} to the
	theory of density fluctuations from slow-roll inflation.
	In \S\ref{sec:desitter-ir} the problem of secular infra-red
	effects in de Sitter correlation functions is outlined, before
	proceeding to the leading-logarithm resummation of these terms
	in \S\ref{sec:DGLAP}.
	
	\subsection{Infra-red divergences in de Sitter space}
	\label{sec:desitter-ir}
	
	The calculation of $n$-point correlation functions during a
	phase of quasi-de Sitter inflation,
	and in particular those for $n \geq 3$,
	has been studied intensively over the last several years.
	The aim is to compute the expectation value, at time $t$,
	of some product of fluctuations $\delta \phi^\alpha$ which begin in
	the vacuum state, for which the appropriate tool is
	Schwinger's formulation of expectation values in terms of
	a closed contour of integration over time \cite{Schwinger:1960qe}.
	Correlation functions in this formalism can be represented
	by the usual Feynman diagrams, except that vertices now exist in
	``$+$'' and ``$-$'' varieties and different propagators are used
	to describe contractions between $(+,+)$, $(+,-)$, $(-,+)$ and $(-,-)$
	vertices. Accounts of this formalism applied to cosmology
	can be found in Refs.~\cite{Weinberg:2005vy,Calzetta:1986ey,
	Jordan:1986ug}. The low order correlation functions therefore
	correspond to
	\begin{equation}
		\parbox{11mm}{
		\begin{fmfgraph*}(30,30)
			\fmfleft{l}
			\fmfright{r}
			\fmf{plain}{l,r}
		\end{fmfgraph*}} ,
		\hspace{3mm}
		\parbox{11mm}{
		\begin{fmfgraph*}(30,30)
			\fmfbottom{b1,b2}
			\fmftop{t}
			\fmf{plain}{b1,v}
			\fmf{plain}{b2,v}
			\fmf{plain}{t,v}
		\end{fmfgraph*}} , \hspace{2mm} \mbox{and}
		\hspace{3mm}\hspace{2mm}
		\parbox{11mm}{
		\begin{fmfgraph*}(30,30)
			\fmfleft{l}
			\fmfright{r}
			\fmftop{t}
			\fmfbottom{b}
			\fmf{plain}{l,v}
			\fmf{plain}{r,v}
			\fmf{plain}{t,v}
			\fmf{plain}{b,v}
		\end{fmfgraph*}}
		\hspace{2mm}
		+
		\hspace{2mm}
		\parbox{18mm}{
		\begin{fmfgraph*}(50,30)
			\fmfleft{l1,l2}
			\fmfright{r1,r2}
			\fmf{plain}{l1,v1}
			\fmf{plain}{l2,v1}
			\fmf{photon,tension=0.5}{v1,v2}
			\fmf{plain}{v2,r1}
			\fmf{plain}{v2,r2}
		\end{fmfgraph*}} .
		\label{eq:correlators}
	\end{equation}
	The first of these is the leading contribution to
	the two-point expectation value;
	the second is the leading contribution to the
	three-point expectation value;
	and the third and fourth contribute comparably to the four-point
	expectation value at leading order.
	The correlation functions have a simple structure in which
	diagrams with $n$ external legs at approximately
	equal 3-momenta $k_\ast$ typically enter proportional
	to $H_\ast^{2(n-1)}$ and a power of $\sqrt{\epsilon_\ast}$ for odd $n$
	\cite{Jarnhus:2007ia}, where $H_\ast \ll 1$ is the Hubble
	parameter at the time the mode $k_\ast$ left the horizon.
	The quantities $H_\ast \sim 10^{-5}$ and $\epsilon_\ast \sim 10^{-2}$,
	whose precise values vary from model to model and depend on the
	inflationary dynamics,
	play the role coupling constants such as the strong coupling $\alpha_s$.
	
	There are several sources of large infra-red effects.
	One source can be found in the second diagram above, which gives
	a contact contribution to the three-point expectation value,
	and is found to contain a term which scales as $\xi_\ast \nplus$
	\cite{Zaldarriaga:2003my,Seery:2008qj,Falk:1992sf},
	where $\xi_\ast \sim \epsilon_\ast$ is
	an auxiliary slow-roll parameter and $\nplus \equiv \ln |k_\ast \eta|$
	is a potentially large logarithm which describes by how many
	e-foldings the mode $k_\ast$ is outside the horizon at
	conformal time $\eta$. (The conformal time is related to
	cosmic time $t$ by a quadrature, $\eta = \int_t^\infty \d t / a(t)$.)
	After a time of order $\xi_\ast^{-1}$ e-folds, this logarithm will
	overwhelm its slow-roll prefactor. It follows that all terms of the
	form $\sim (\epsilon \nplus)^n$ become comparably large,
	after which any such
	expansion will require resummation \cite{Gong:2001he}.
	This was interpreted in terms of a coherent time
	evolution of the background field by Zaldarriaga
	\cite{Zaldarriaga:2003my},
	which can be understood very simply.
	Once a fluctuation has passed outside the horizon its time evolution
	becomes
	approximately classical, even accounting for
	the inclusion of non-linear effects \cite{Lyth:2006qz}.
	To leading order in gradients, the fluctuation simply amounts to shifting
	the background field by some amount and must therefore evolve
	coherently with it.
	More generally, each of the diagrams
	in~\eref{eq:correlators} is typically calculated
	by working to leading order in quantities of order $\epsilon$,
	and higher powers in the expansion should be expected to
	be accompanied by powers of $\nplus$ which will invalidate the use
	of perturbation theory after $\nplus \sim \epsilon_\ast^{-1}$ e-folds
	\cite{Gong:2001he,Gong:2002cx}.
	
	Loop diagrams contribute different infra-red effects.
	These are systematic corrections to each of the $n$-point graphs
	which account for processes by which an external particle
	radiates into new quanta, which later coalesce and exit the diagram
	by a different external leg.
	Certain corrections to the two- and three-point correlation functions
	are now known
	\cite{Weinberg:2005vy,Weinberg:2006ac,Sloth:2006az,Sloth:2006nu,
	vanderMeulen:2007ah,Seery:2007we,Dimastrogiovanni:2008af,
	Seery:2008ms,Adshead:2008gk}, of the form
	\begin{center}
		\vspace{5mm}
		\parbox{15mm}{
		\begin{fmfgraph*}(40,30)
			\fmfleft{l}
			\fmfright{r}
			\fmf{plain}{l,v}
			\fmf{plain,right,tension=0.5}{v,v}
			\fmf{plain}{v,r}
		\end{fmfgraph*}}
		\hspace{2mm}
		and
		\hspace{2mm}
		\parbox{15mm}{
		\begin{fmfgraph*}(40,30)
			\fmfleft{l}
			\fmfright{r}
			\fmf{plain}{l,v}
			\fmf{photon,right,tension=0.5}{v,v}
			\fmf{plain}{v,r}
		\end{fmfgraph*}} ,
	\end{center}
	where the diagram on the left describes a scalar loop and the
	diagram on the right a circulating graviton.
	For each loop, these diagrams typically enter suppressed by one power
	of $H_\ast^2$ and involve one unconstrained integral over momentum
	space. In some circumstances this integral
	reduces to $\int \d \ln k$ and---neglecting ultra-violet effects,
	which can be accommodated separately
	\cite{vanderMeulen:2007ah,Seery:2008ms}---%
	where such divergences overlap
	the loop expansion is effectively a power series in
	$(H_\ast^2 \ln k_\ast/k_0)^n$, where $k_0$ is an infra-red cutoff
	of order the comoving Hubble length at the onset of inflation.
	The duration of the inflationary era when the scale $k_\ast$
	corresponds to the Hubble scale is therefore
	$\nminus \equiv \ln k_\ast/k_0$ e-folds,
	and
	terms of this sort will spoil the convergence of perturbation theory
	at horizon exit for $k_\ast$-scale modes
	whenever $\nminus \sim H_\ast^{-2}$.
	More precise estimates can be found in
	Refs.~\cite{Sloth:2006az,Sloth:2006nu,vanderMeulen:2007ah,
	Seery:2007we}. The quantities $\nplus$ and $\nminus$ therefore play
	different but complementary roles.
	
	How are we to understand the meaning of such infra-red
	divergences? It is clear that their structure is similar to
	those encountered in computing corrections to the parton
	distribution functions of hadrons, with $\ln k_\ast$
	playing the role of the probe scale $\ln Q^2$; this scale
	appears with different hierarchies in $\nminus$ and $\nplus$,
	corresponding to different ``resolutions'' for the ingoing and
	outgoing particles in the hadron picture.
	It is reasonable
	to suppose that the explanation of these large effects is similar
	to the large infra-red effects of QCD, with $\nminus$ effects
	resummed into distribution functions and $\nplus$ effects
	resummed into fragmentation functions. This leads to a highly
	intuitive picture.
	Indeed, in the case of hadrons, resummation
	entails considering dressed diagrams
	with extra soft particles in the initial or final states.
	In a box of de Sitter space this would correspond to dressing the
	diagrams of~\eref{eq:correlators} with soft quanta belonging to
	all fields which are light during inflation.
	However, on much smaller scales these soft quanta are tantamount
	to no more than a redefinition of the background fields,
	and it is easy to guess that summing over them corresponds to
	averaging over the different spacetime backgrounds which
	could have been produced by quantum fluctuations
	during the evolution of the box.
		
	\subsection{The inflationary DGLAP equation}
	\label{sec:DGLAP}
	
	We are now in a position to discuss an analogue of the DGLAP equation
	for de Sitter space.
	In this section, we focus on the $\nminus$ divergences which would lead
	to evolution of the parton distribution functions, before
	moving on to the issue of $\nplus$ divergences in \S\ref{sec:predictions}.
	
	\paragraph{Light-cone coordinates.}
	In the case of hadron physics, the fields
	whose correlation functions exhibit secular infra-red effects
	are spin-1/2
	fermions interacting by mediation of spin-1 gauge bosons.
	Let us imagine a deep inelastic scattering event which takes place
	in a $(d+1)$-dimensional spacetime, with coordinates $(t,\vect{x},z)$
	where $t$ is coordinate time,
	$z$ labels a particular spatial dimension,
	and $\vect{x}$ is a $(d-1)$-dimensional vector.
	There is a preferred axis associated with the collision of probe and
	hadron, commonly taken to be the $z$ axis, which permits the introduction
	of light-cone coordinates $x^\pm$ satisfying
	$x^\pm \equiv (t \pm z)/\sqrt{2}$.
	If one shifts to the frame in which the target hadron has infinite
	momentum longitudinal to the collision axis, the colour field of the
	hadron is is shrunk to have support only at $x^- = 0$.
	At the same time,
	the infinite time dilation observed by the probe relative
	to the rapidly moving hadron implies that the colour field is
	independent of $x^+$.
	For each species $i$, we introduce a field $x_i(\vect{x})$
	(which for simplicity I will refer to as a Bjorken field)
	representing the amplitude for a parton of species $i$ to manifest
	itself with momentum fraction $x_i$ at location $\vect{x}$.
	These fields
	describe spatial correlations among the partons,
	and are determined by the loop expansion of correlation functions
	in the underlying gauge theory.
	After coarse-graining the modulus-square of the
	wavefunctional $\Psi_i[x_i]$ 
	associated with the $x_i$ over
	$\vect{x}$, we can recover the parton distribution function $f_i(x)$.
		
	Consider a patch of de Sitter spacetime sourced by the potential energy
	of some number of scalar fields $\phi^\alpha$ which are
	labelled by Greek indices $\{ \alpha, \beta, \ldots \}$.
	We identify these fields with the Bjorken fields, and
	associate the spatial
	dependence they carry with the coordinates $\vect{x}$ which were
	transverse to the light-cone coordinates $x^\pm$.
	The renormalization group coordinate $x^+$ is associated with
	time evolution.
	It may also be necessary to introduce extra
	Bjorken fields to describe the physical polarizations of
	other light particles such as gravitons or gauge bosons.
	There will be a coarse-grained parton
	distribution function $f_\alpha(\varphi)$
	for each of these---in principle obtained by coarse-graining over
	the appropriate wavefunctional, as above---or one can work with a joint
	parton distribution function $f(\varphi^\alpha)$ which
	describes all the parton species
	simultaneously. I will choose to focus on
	the latter, since it contains more information and is required
	for applications to inflation.
	The parton distribution functions will suffice for resummation of
	some large infra-red effects, although to deal with all large terms
	it will later be necessary to introduce analogues of the
	fragmentation functions.

	\paragraph{Initial conditions.}
	In analogy with the DGLAP evolution for
	an isolated electron, we can imagine an initial
	Hubble-sized
	region characterized by a set of uniform scalar expectation values
	$\bar{\phi}^\alpha$. This initial
	box can be compared to an initial parton---%
	such as an electron---at momentum fraction $x=1$
	and yields an initial condition analogous to the electron,
	$f(\varphi^\alpha) = \prod_\alpha \delta(\varphi^\alpha -
	\bar{\phi}^\alpha)$ at the initial time.
	In the same way that the details of the hadron wavefunction
	depend on unknown soft QCD effects,
	the wavefunction for the de Sitter state depends on the unknown details
	of quantum gravity which are likewise presently incalculable.
	
	If we probe the box at the original Hubble
	scale we must find that the scalars take the values
	$\bar{\phi}^\alpha$ everywhere. The analogue of strict Bjorken scaling
	would correspond to these initial values remaining fixed
	everywhere in the box, no matter at which scale we probe it.
	However,
	as with the case of parton evolution within a hadron, strict
	Bjorken scaling is violated---in this case, by terms such as
	$\xi_\ast \nminus$ and $H_\ast^2 \nminus$.
	The former type account for simple
	coherent time evolution within the box, whereas the latter type account
	for the possibility of an emission cascade analogous to
	Fig.~\ref{fig:braking-radiation}.
	In the case of a de Sitter box, this
	emission cascade corresponds to gravitational particle production
	on scales much larger than the probe.
	
	In this section we are only aiming to study the distribution
	of Bjorken variables encountered by a mode leaving the horizon
	at an arbitrary point during inflation, which will account for large
	$\nminus$-type terms and corresponds to predicting the distribution
	of partons encountered by an incoming hard particle.
	To account for all large infra-red terms one must also accommodate
	the process by which the debris from a hard scattering event
	is unwound into final state particles, which absorbs large
	$\nplus$-type terms. The analysis of these terms is deferred to
	\S\ref{sec:predictions}.

	\paragraph{Splitting functions.}
	The first step in writing a DGLAP equation for
	the parton distribution functions
	is to identify the relevant splitting functions,
	which give the probability for some parton characterized by
	the Bjorken variables $\varphi^\alpha$ to split into
	a parton characterized by
	$\varphi^\alpha + \delta \varphi^\alpha$ together with other soft quanta.
	In the hadron case one performs this calculation using the formalism of
	cut diagrams, in which one effectively calculates
	$x^-$-ordered correlation functions
	with lines crossing the surface $x^- = 0$ put on-shell.
	This bears a strong formal relationship with the Schwinger formalism
	employed in de Sitter calculations.
	
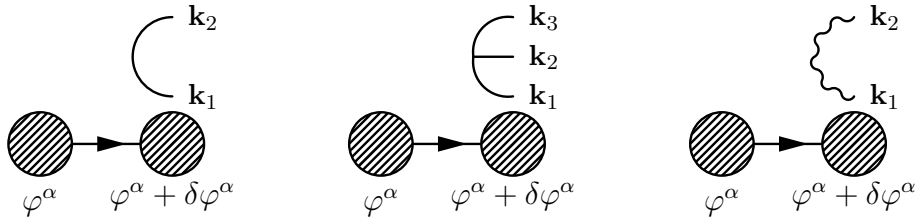
\begin{figure}
		\begin{center}
			\vspace{2mm}
			\begin{fmfgraph*}(50,60)
				\fmfleft{l}
				\fmfright{r,a,b}
				\fmf{fermion}{l,r}
				\fmf{plain,left}{a,b}
				\fmfforce{(0,0.2h)}{l}
				\fmfforce{(w,0.2h)}{r}
				\fmfforce{(w,h)}{b}
				\fmfforce{(w,0.5h)}{a}
				\fmffreeze
				\fmfv{decoration.shape=circle,decoration.filled=shaded,
				      decoration.size=0.4h}{l,r}
				\fmfv{label=$\small\varphi^\alpha$,
				      label.angle=-90,label.dist=6mm}{l}
				\fmfv{label=$\small\varphi^\alpha+\delta\varphi^\alpha$,
				      label.angle=-90,label.dist=5mm}{r}
				\fmfv{label=$\small\vect{k}_1$,label.angle=0}{a}
				\fmfv{label=$\small\vect{k}_2$,label.angle=0}{b}
			\end{fmfgraph*}
			\hspace{25mm}
			\begin{fmfgraph*}(50,60)
				\fmfleft{l}
				\fmfright{r,a,b,c}
				\fmf{fermion}{l,r}
				\fmf{plain,left=0.5}{a,v}
				\fmf{plain,left=0.5}{v,b}
				\fmf{plain}{v,c}
				\fmfforce{(0,0.2h)}{l}
				\fmfforce{(w,0.2h)}{r}
				\fmfforce{(w,h)}{b}
				\fmfforce{(w,0.5h)}{a}
				\fmfforce{(w,0.75h)}{c}
				\fmfforce{(w-0.25h,0.75h)}{v}
				\fmffreeze
				\fmfv{decoration.shape=circle,decoration.filled=shaded,
				      decoration.size=0.4h}{l,r}
				\fmfv{label=$\small\varphi^\alpha$,
				      label.angle=-90,label.dist=6mm}{l}
				\fmfv{label=$\small\varphi^\alpha+\delta\varphi^\alpha$,
				      label.angle=-90,label.dist=5mm}{r}
				\fmfv{label=$\small\vect{k}_1$,label.angle=0}{a}
				\fmfv{label=$\small\vect{k}_2$,label.angle=0}{c}
				\fmfv{label=$\small\vect{k}_3$,label.angle=0}{b}
			\end{fmfgraph*}
			\hspace{25mm}
			\begin{fmfgraph*}(50,60)
				\fmfleft{l}
				\fmfright{r,a,b}
				\fmf{fermion}{l,r}
				\fmf{photon,left}{a,b}
				\fmfforce{(0,0.2h)}{l}
				\fmfforce{(w,0.2h)}{r}
				\fmfforce{(w,h)}{b}
				\fmfforce{(w,0.5h)}{a}
				\fmffreeze
				\fmfv{decoration.shape=circle,decoration.filled=shaded,
				      decoration.size=0.4h}{l,r}
				\fmfv{label=$\small\varphi^\alpha$,
				      label.angle=-90,label.dist=6mm}{l}
				\fmfv{label=$\small\varphi^\alpha+\delta\varphi^\alpha$,
				      label.angle=-90,label.dist=5mm}{r}
				\fmfv{label=$\small\vect{k}_1$,label.angle=0}{a}
				\fmfv{label=$\small\vect{k}_2$,label.angle=0}{b}
			\end{fmfgraph*}
		\end{center}
		\caption{\label{fig:ds-splittings}The analogue of
		Altarelli--Parisi splitting functions for de Sitter space.
		A de Sitter parton, represent by the hatched region,
		evolves in time from left to right and
		radiates soft quanta which materialize in the final state.
		These radiated quanta are fluctuations which are
		instantaneously drawn over the de Sitter horizon and
		classicalize.
		The solid lines represent scalar particles whereas the wavy
		line represents gravitons;
		the diagram with three scalar particles in the final state
		represents the leading non-Gaussian correction,
		although this turns out not to contribute in a leading-logarithm
		approximation.
		In principle, radiation into
		\emph{any} light states is permitted.}
	\end{figure}

	One can represent these processes by the diagrams of
	Fig.~\ref{fig:ds-splittings}, in which
	a non-perturbative de Sitter parton represented
	by the hatched area evolves in time from left to right.
	As it does so, it can radiate into any quanta which are light
	on the Hubble scale. Radiation of a single particle is forbidden by
	the tadpole condition, so in the leading diagram two quanta
	materialize in the ``final state.''%
		\footnote{This terminology is intuitive but imprecise,
		because in the de Sitter case we \emph{cannot} interpret
		a correlation function dressed by soft quanta to
		represent a shift in the initial or final state experienced
		by some hard subprocess.
		In the de Sitter case, the state experienced by the hard
		subprocess is
		produced by gravitational expansion of the
		Minkowski vacuum on small scales, and indeed may contain an
		indefinite number of particles. What we are measuring are the
		\emph{correlations} among certain configurations of particles in
		this state. Nevertheless, I will use this terminology freely
		because it is familiar from the hadron case.}
	However, in principle any number of particles may be produced,
	corresponding to inclusion of any of the diagrams
	described by~\eref{eq:correlators} or their
	higher $n$ generalizations;%
		\footnote{Recall that in~\eref{eq:correlators}, all external
		legs are evaluated at the same time, and in the time-ordered
		diagrams of Fig.~\ref{fig:ds-splittings} they therefore
		appear with all legs on the right-hand side.
		One can think of the Schwinger-formalism
		diagrams as being rather more like instantons,
		in comparison with Feynman-formalism diagrams where
		particles pass through the diagram in a definite direction.}
	for example, the leading
	non-Gaussian correction corresponds to a final state in which
	three quanta are present.
	In practice we will see that---as in the hadron case---%
	final states with more than
	the minimum number of radiated quanta do not contribute in the
	leading-logarithm approximation.
	
	In de Sitter space there is no analogue of an S-matrix, so a
	different rule is needed to convert correlation functions into
	splitting probabilities. The appropriate prescription can be found by
	reconstructing the wavefunction on configurations of the
	background fields, leading to \cite{Seery:2006wk}
	\begin{eqnarray}
		\fl\nonumber
		P[\varphi^\alpha(\vect{x})] \propto
		\int \prod_\alpha [\d \eta_\alpha]
		\\ \nonumber
		\mbox{} \times
		\exp\left(
			\sum_{n=0}^\infty \frac{\im^n}{n!}
			\int \prod_{j = 1}^{n} \d^3 x_j \;
			\eta_\alpha(\vect{x}_1) \cdots \eta_\beta(\vect{x}_n)
			\langle \delta \phi^\alpha(\vect{x}_1) \cdots
				\delta \phi^\beta(\vect{x}_n) \rangle
		\right)
		\\
		\mbox{} \times
		\exp\left( - \im \int \d^3 x \;\eta_\alpha(\vect{x})
			\varphi^\alpha(\vect{x})
		\right) ,
		\label{eq:prob-measure}
	\end{eqnarray}
	where $P[\varphi^\alpha]$ is a functional
	measuring the relative probabilities
	of the field configurations
	$\varphi^\alpha(\vect{x})$ and
	$[\d \eta]$ denotes functional integration over the field
	$\eta(\vect{x})$.
	Note that Eq.~\eref{eq:prob-measure} contains another source of
	infra-red divergences, in the form of each integral over
	$\vect{x}_j$.
	
	\paragraph{Resummation of large logarithms.}
	We would like to obtain the probability for a region within the
	de Sitter box with approximately constant background
	scalar expectation values on a scale $k^{-1}$
	to evolve into a slightly smaller
	region, over which the scalars may take different vevs.
	This will be the de Sitter
	analogue of the Altarelli--Parisi splitting function.
	Suppose that the de Sitter region evolves for a short time,
	amounting to roughly
	$\delta N \approx \delta \ln k$ e-folds, during which a
	fluctuation is imprinted in the each scalar field---or, more
	generally, each light degree of freedom---%
	over a narrow range
	of scales $\delta k$.
	The Fourier transform of a constant field configuration
	$\varphi^\alpha(\vect{x}) \approx \sigma^\alpha$ on the
	scale $(k + \delta k)^{-1}$ is roughly
	\begin{equation}
		\varphi^\alpha(\vect{k})
		\approx \frac{2\pi^2}{k^3} \frac{\sigma^\alpha}{\delta \ln k} ,
	\end{equation}
	to leading order in $\delta k$.
	It follows that the probability for quantum fluctuations to
	generate a region of size $(k + \delta k)^{-1}$ in which the
	vev of a scalar species $\alpha$ is offset by an amount
	$\sigma^\alpha$ satisfies
	\begin{eqnarray}
		\fl\nonumber
		P(\sigma^\alpha) \propto
		\int \prod_\beta \d \eta_\beta
		\left( 1 - \frac{1}{2} \eta_\gamma \eta_\delta
			\Ps_\ast^{\gamma \delta}(k)
			\delta \ln k + \Or(\delta \ln k)^2 \right)
		\exp(- \im \eta_\kappa \sigma^\kappa )
		\\
		\mbox{} =
		\prod_\delta \delta(\sigma^\delta)
		+ \frac{1}{2} (\delta \ln k) \Ps_\ast^{\beta \gamma}
			\frac{\partial}{\partial \sigma^\beta}
			\frac{\partial}{\partial \sigma^\gamma}
			\prod_\delta \delta(\sigma^\delta) + \Or(\delta \ln k)^2,
		\label{eq:splitting}
	\end{eqnarray}
	where the $\eta$ integrals in the first line
	are now one-dimensional and run from
	$-\infty$ to $\infty$. The quantity
	$\Ps_\ast(k)$ is the so-called
	``dimensionless'' power spectrum imprinted in the range
	$\delta k$, and is defined by the rule
	\begin{equation}
		\langle \delta \phi^\alpha(\vect{k}_1)
			\delta \phi^\beta(\vect{k}_2) \rangle_\ast
		= (2\pi)^3 \delta(\vect{k}_1 + \vect{k}_2)
			\frac{2\pi^2}{k^3} \Ps^{\alpha \beta}_\ast(k) ,
	\end{equation}
	where $|\vect{k}_1| = |\vect{k}_2| = k$.
	In principle the coefficient of the leading $\delta$-function
	could be modified by a term of order $\delta \ln k$ after
	correctly normalizing this distribution, but since $\Ps_\ast$
	is independent of the $\sigma^\alpha$ this does not occur
	and Eq.~\eref{eq:splitting} is correct as it stands.
	Eq.~\eref{eq:splitting}
	has the same interpretation as the hadronic
	Altarelli--Parisi function, Eq.~\eref{eq:splitting-function}.
	If $\delta \ln k = 0$, then there is no phase space for
	splitting to occur and the parton must remain with the same
	scalar vacuum expectation values. This is described by the leading
	$\delta$-function, which enforces $\sigma^\alpha = 0$.
	For finite $\delta \ln k$ there is
	a small phase space for the vacuum expectation values to shift,
	which is described by the second term, proportional to
	$(\delta \ln k) \delta''(\sigma)$.
	In principle this series could be carried to higher orders in
	$\delta \ln k$, generating a Kramers--Moyal expansion.
	
	Comparison of Eqs.~\eref{eq:splitting-function} and~\eref{eq:splitting}
	highlights an interesting difference between the hadron and de Sitter
	cases. When we calculate splitting functions for partons, it is
	possible to work non-perturbatively in the Bjorken variable, $x$.
	In the de Sitter case, our answer is not only perturbative in the
	``coupling constant'' $H_\ast^2$, but is \emph{also}
	an expansion in the shifts of the Bjorken variables.
	In principle this could lead to significant difficulties.
	Had we directly integrated out the auxiliary variable $\eta^\alpha$
	in Eq.~\eref{eq:splitting}, we would have obtained a Gaussian
	distribution in $\sigma^\alpha$ valid only for
	$|\sigma^\alpha|^2 \lesssim H_\ast^2 \delta \ln k$ (for each
	$\alpha$).
	To construct the master equation, however, we must integrate over
	the entire range of the Bjorken variable and for the vast
	majority of this range our perturbative formula is invalid.
	One must then enquire why we should imagine the final
	formula in Eq.~\eref{eq:splitting} to be correct. The reason is that
	we expect the random walk executed by the Bjorken
	variables to be almost-local, in the sense that transitions only occur
	between approximately neighbouring states. For such transitions
	Eq.~\eref{eq:splitting} gives an accurate representation. It is
	not important that it may give a very poor approximation for non-local
	transitions where the Bjorken variables jump by a macroscopic
	amount in a single step, because such transitions are exponentially
	unlikely.
	Eq.~\eref{eq:splitting} can be thought of
	as a formal, analytic regularization which captures this concept of
	locality in the random walk. It will lead to a solution with the
	character of a diffusion process.
	
	It is important that neither $\Ps$ nor any of the correlation
	functions in Eq.~\eref{eq:prob-measure}
	can contain large infra-red logarithms, because they are all
	evaluated within a box of size $\delta \ln k$
	e-folds. The potentially large terms
	$H_\ast^2 \nminus$ and $\epsilon_\ast \nminus$ are therefore
	all very small, and indeed will disappear in the continuum limit
	where we take $\delta \ln k \rightarrow 0$.
	
	Let us return to the DGLAP equation.
	The probability that, when probed at a scale $(k + \delta k)^{-1}$,
	a region within the de Sitter box resolves to approximately
	constant scalar expectation values $\varphi^\alpha$ can be written
	\begin{equation}
		\fl
		f(\varphi^\alpha; N + \delta \ln k) =
		\int_{-\infty}^\infty
		\Big(
			\prod_{\beta, \gamma} \d \rho^\beta \;
			\d \sigma^\gamma
		\Big) \;
		f(\rho^\alpha; N) P(\sigma^\alpha)
		\delta(\rho^\alpha + \sigma^\alpha + \delta\varphi^\alpha -
		\varphi^\alpha) ,
		\label{eq:desitter-master}
	\end{equation}
	where $\delta \varphi^\alpha \propto \delta \ln k$
	is a function of the $\rho^\alpha$,
	accounting for the coherent time evolution
	of the background fields, which is the de Sitter analogue of
	renormalization-group evolution sourced by self-energy diagrams
	(eg. see pp. 28--30 of Ref.~\cite{Weigert:2005us}).
	We will assume that the time evolution is given to a good
	approximation by the separate universe formula
	\begin{equation}
		3 H \dot{\varphi}^\alpha = - \delta^{\alpha \beta} V_{,\beta} ,
		\label{eq:slow-roll}
	\end{equation}
	where $V = V(\phi^\alpha)$ is an arbitrary potential
	supporting a phase of slow-roll evolution.
	Eq.~\eref{eq:desitter-master} is the de Sitter master equation,
	which is an exact analogue of the hadron master equation,
	Eq.~\eref{eq:altarelli-parisi}.
	As in the case of hadrons, it is of Chapman--Kolmogorov type and
	describes a Markov process in which the
	Bjorken variable executes a
	random walk as the probe scale is varied.
	When interpreted as a collisional Boltzmann equation,
	Eq.~\eref{eq:desitter-master}
	can also be related to Polyakov's discussion of stability
	in de Sitter space in the presence of particle production
	(compare {\S}5 of Ref.~\cite{Polyakov:2007mm}).
	
	In the limit $\delta \ln k \rightarrow 0$ one obtains
	its continuum counterpart, which exactly generalizes the
	DGLAP equation,
	\begin{equation}
		\frac{\partial f}{\partial \ln k}
		= \frac{1}{2}
			\partial_\alpha \partial_\beta ( \Ps^{\alpha \beta} f )
			- \partial_\alpha \left( f
			\frac{\d \varphi^\alpha}{\d \ln k}
			\right) ,
		\label{eq:desitter-DGLAP}
	\end{equation}
	where $\partial_\alpha \equiv \partial / \partial \varphi^\alpha$,
	and $\Ps^{\alpha\beta}$ is a function of the $\varphi^\alpha$.
	Eq.~\eref{eq:desitter-DGLAP} can immediately be recognized
	as Starobinsky's diffusion (``Fokker--Planck'') equation---%
	originally obtained by interpreting $f$ as a probability density
	associated with solutions to a Langevin equation.
	The generalization to multiple-fields was studied
	in Refs.~\cite{Mollerach:1990zf,GarciaBellido:1995kc,Amendola:1993pp}.
	Specializing to the case of
	single-field inflation, it follows that
	\begin{equation}
		\frac{1}{H} \frac{\partial f}{\partial t}
		= \frac{\partial}{\partial \varphi}
			\left( \frac{V' f}{3H^2} \right)
			+ \frac{H^2}{8\pi^2} \frac{\partial^2 f}{\partial \varphi^2} ,
		\label{eq:starobinsky-fokker}
	\end{equation}
	which is the form of this equation obtained
	in Ref.~\cite{Starobinsky:1986fx} provided $H$ is taken to be constant.
	If desired, one could now reverse the argument of
	\ref{sec:fokkerplanck} and obtain a Langevin equation for each of
	the fields $\phi^\alpha$, in terms of which the analysis may
	simplify in practice \cite{Martin:2005ir}. However,
	the physical content of the
	Langevin and Fokker--Planck equations is the same and moreover one can
	move freely from one representation to the other,
	so it is also
	possible to think of the Langevin equation as a de Sitter
	renormalization group equation. In this version, the importance of the
	slow-roll approximation in reducing the de Sitter evolution
	to a renormalization group flow is clear: this description is only
	valid when the $\ddot{\varphi}$ term in Eq.~\eref{eq:slow-roll}
	is irrelevant.
		
	\subsection{Leading logarithms}
	
	Eq.~\eref{eq:desitter-DGLAP} is not exact unless $\Ps_\ast$ can be
	evaluated precisely. In most applications, it will only be possible
	to obtain $\Ps_\ast$ perturbatively in
	terms of order $H_\ast^2$ or $\epsilon_\ast$. In these cases,
	Eq.~\eref{eq:desitter-DGLAP} should give a correct resummation
	of all terms of the form $(H_\ast^2 \nminus)^n$
	and $(\epsilon_\ast \nminus)^n$, but will not account
	for terms which are suppressed compared to these by extra powers of
	$H_\ast^2$ or $\epsilon_\ast$. Therefore,
	one should think of the
	Starobinsky--DGLAP equation~\eref{eq:desitter-DGLAP}
	as a leading-logarithm resummation. The possibility of this type of
	resummation was raised by Weinberg
	\cite{Weinberg:2006ac}.
	
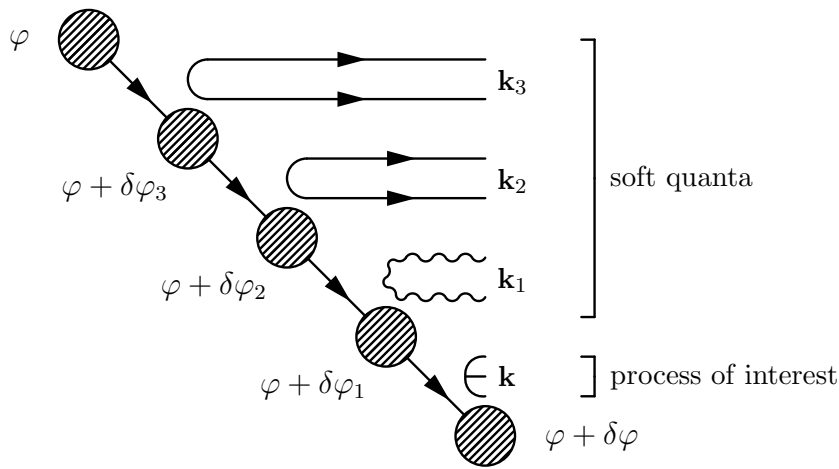
\begin{figure}
		\vspace{5mm}
		\begin{center}
			\begin{fmfgraph*}(150,150)
				\fmfleft{l}
				\fmfright{a1,a2,b1,b2,c1,c2,d1,d2,d3,r}
				\fmf{fermion}{l,v1,v2,v3,r}
				\fmf{fermion}{w1,a1}
				\fmf{plain,right}{w1,w2}
				\fmf{fermion}{w2,a2}
				\fmf{fermion}{x1,b1}
				\fmf{plain,right}{x1,x2}
				\fmf{fermion}{x2,b2}
				\fmf{photon}{c1,y1}
				\fmf{photon,right}{y1,y2}
				\fmf{photon}{y2,c2}
				\fmf{plain,right=0.5}{d1,z}
				\fmf{plain,right=0.5}{z,d2}
				\fmf{plain}{z,d3}
				\fmfforce{(0,h)}{l}
				\fmfforce{(0.25w,0.75h)}{v1}
				\fmfforce{(0.5w,0.5h)}{v2}
				\fmfforce{(0.75w,0.25h)}{v3}
				\fmfforce{(w,0)}{r}
				\fmfforce{(w,0.85h)}{a2}
				\fmfforce{(w,0.95h)}{a1}
				\fmfforce{(0.3w,0.85h)}{w2}
				\fmfforce{(0.3w,0.95h)}{w1}
				\fmfforce{(w,0.6h)}{b2}
				\fmfforce{(w,0.7h)}{b1}
				\fmfforce{(0.55w,0.6h)}{x2}
				\fmfforce{(0.55w,0.7h)}{x1}
				\fmfforce{(w,0.35h)}{c2}
				\fmfforce{(w,0.45h)}{c1}
				\fmfforce{(0.8w,0.35h)}{y2}
				\fmfforce{(0.8w,0.45h)}{y1}
				\fmfforce{(w,0.1h)}{d2}
				\fmfforce{(w,0.2h)}{d1}
				\fmfforce{(w,0.15h)}{d3}
				\fmfforce{(0.95w,0.15h)}{z}
				\fmffreeze
				\fmfv{decoration.shape=circle,decoration.filled=shaded,
				      decoration.size=0.15h,
				      label=$\small\varphi$,label.angle=180,
				      label.dist=0.15h}{l}
				\fmfv{decoration.shape=circle,decoration.filled=shaded,
				      decoration.size=0.15h,
				      label=$\small\varphi + \delta\varphi_3$,
				      label.angle=-120,label.dist=0.1h}{v1}
				\fmfv{decoration.shape=circle,decoration.filled=shaded,
				      decoration.size=0.15h,
				      label=$\small\varphi + \delta\varphi_2$,
				      label.angle=-120,label.dist=0.1h}{v2}
				\fmfv{decoration.shape=circle,decoration.filled=shaded,
				      decoration.size=0.15h,
				      label=$\small\varphi + \delta\varphi_1$,
				      label.angle=-120,label.dist=0.1h}{v3}
				\fmfv{decoration.shape=circle,decoration.filled=shaded,
				      decoration.size=0.15h,
				      label=$\small\varphi + \delta\varphi$,
				      label.angle=0,label.dist=0.15h}{r}
				\fmfv{label=$\small\vect{k}_3$,label.angle=-40}{a1}
				\fmfv{label=$\small\vect{k}_2$,label.angle=-40}{b1}
				\fmfv{label=$\small\vect{k}_1$,label.angle=-40}{c1}
				\fmfv{label=$\small\vect{k}$,label.angle=-35}{d1}
			\end{fmfgraph*}
			\hspace{1cm}
			\begin{fmfgraph*}(50,150)
				\fmfleft{l1,l2,a1,b1}
				\fmf{plain}{l1,w1,w2,w3,l2}
				\fmfforce{(0,h)}{l1}
				\fmfforce{(0.1w,h)}{w1}
				\fmfforce{(0.1w,0.65h)}{w2}
				\fmfforce{(0.1w,0.3h)}{w3}
				\fmfforce{(0,0.3h)}{l2}
				\fmf{plain}{a1,x1,x2,x3,a2}
				\fmfforce{(0,0.2h)}{a1}
				\fmfforce{(0.1w,0.2h)}{x1}
				\fmfforce{(0.1w,0.15h)}{x2}
				\fmfforce{(0.1w,0.1h)}{x3}
				\fmfforce{(0,0.1h)}{a2}
				\fmffreeze
				\fmfv{label=$\mbox{\small soft quanta}$,label.angle=0}{w2}
				\fmfv{label=$\mbox{\small process of interest}$,
				      label.angle=0}{x2}
			\end{fmfgraph*}
		\end{center}
		\caption{\label{fig:desitter-cascade}The spacetime interpretation
		of a leading-logarithm resummation in $H_\ast^2$.}
	\end{figure}

	One can formulate a spacetime picture
	(see Fig.~\ref{fig:desitter-cascade}) of the leading--logarithm
	resummation by analogy with the emission cascade of
	Fig.~\ref{fig:braking-radiation}.
	In this picture, a de Sitter parton evolves as one changes the scale
	on which it is probed by radiating soft quanta belonging to
	all light fields. The region of phase space corresponding to
	terms of leading-logarithm order arises from splitting into a single
	pair of quanta where the emission chain satisfies a
	strong ordering criterion,
	\begin{equation}
		k_0 \ll \cdots \ll k_3 \ll k_2 \ll k_1 \ll k ,
		\label{eq:desitter-strongorder}
	\end{equation}
	where $k$ is the scale at which one wishes to compute a correlation
	function of fluctuations, and $k_0$ is the scale of the initial
	de Sitter box. These soft quanta from intermediate splittings
	propagate to the right of the diagram, where they dress the
	process of interest---depicted in Fig.~\ref{fig:desitter-cascade}
	as a correlation function with three quanta in the final state,
	although any process of interest can be substituted at this stage---%
	with extra final-state particles.
	Eq.~\eref{eq:desitter-strongorder} is exactly
	analogous to the hadron strong ordering
	condition~\eref{eq:strong-order}, and has the same interpretation.
	Although other configurations of emission cascade exist in which
	the de Sitter parton splits into three or more quanta at an intermediate
	stage, these splittings cost an extra factor of $H_\ast^2$
	for which there is no compensating large logarithm. These terms
	therefore contribute \emph{formally} at next-to-leading logarithmic
	order or below and, if desired,
	could be included to find the next-order corrections to
	the DGLAP equation.
	On the other hand,
	this line of reasoning immediately
	suggests that the Starobinsky--DGLAP equation no longer gives a useful
	resummation in the strongly coupled
	region where $H_\ast$ approaches the
	Planck mass $\Mp$.
	In this regime there is apparently nothing
	to suppress splitting into an arbitrary number of quanta
	at each step in the DGLAP evolution and the situation rapidly
	escapes beyond any sort of perturbative control.
	
	\section{Predictions for inflationary correlation functions}
	\label{sec:predictions}

	The DGLAP framework provides a framework
	within which to understand the question of infra-red divergences
	in inflationary correlation functions, and
	in addition---at least in the context of the leading-logarithm
	approximation---it
	provides a physical picture of their effects.
	The de Sitter master equation, Eq.~\eref{eq:desitter-master},
	only requires correlation functions to be computed within
	boxes of size $\delta N \sim \delta \ln k$ e-folds,
	cutting off the possibility of large infra-red logarithms.
	However, it is still necessary to give a rule which allows this formalism
	to be used for the purpose of obtaining predictions which can be
	compared to observation.
	
	\subsection{The core hard subprocess}
	
	By analogy with the hadronic case, we expect to identify three
	distinct phases in the calculation of an observable quantity.
	In the first phase, a radiative cascade described by DGLAP evolution
	brings the initial conditions
	(which are set at some low infra-red scale)
	up to an energy where the core hard
	subprocess can be calculated perturbatively. The details of
	this hard subprocess itself
	constitute the second phase. Finally, in the third phase
	(described by fragmentation functions), the radiative cascade is
	unwound to obtain an observable low-energy final state.
	In the case of QCD, this final state
	is composed of well-separated showers of
	colourless, hadronized degrees of
	freedom.
	
	The DGLAP evolution appropriate to the first phase was studied in
	\S\ref{sec:DGLAP}.
	We must also expect to find de Sitter analogues of the second
	and third phases:
	resummation of large logarithms, however necessary, is not usually
	sufficient to capture every process of interest.
	Let us suppose that the distribution of Bjorken variables within
	the original de Sitter box has been evolved to some
	scale $k_F$ according to the DGLAP prescription.
	Quantities which can be observed in the temperature anisotropy of the
	CMB, or the distribution of galaxies, are controlled by 
	correlations between modes which exit the horizon over a very short
	space of e-foldings---typically $\Delta N \lesssim 10$, but
	at present almost certainly $\Delta N < 10^2$.
	Note, however, that the range of observable
	e-foldings grows with time, and is therefore not subject to any
	restriction as a matter of principle.
	We can imagine choosing $k_F$ to be characteristic of the
	horizon size at the epoch when scales of interest are leaving the
	Hubble radius during inflation.
	The detailed correlations among the modes which leave over the
	next $\Delta N \lesssim 10$ e-folds---controlled by calculation
	within a small box---constitute the analogue of
	the hard subprocess in deep inelastic scattering,
	in which large logarithms were under control
	in virtue of the large momentum transfer, $Q^2$,
	in Eq.~\eref{eq:parton-cross-section}.

	The hard subprocess is characterized by time- and lengthscales which
	are short compared with those which lead to significant enhancement
	of infra-red terms. 
	Accordingly, the subprocess
	should be calculated using fixed-order perturbation
	theory to whatever precision is necessary.
	For example: if the aim is to
	study non-Gaussian features in the statistics of the CMB,
	then to compute the bispectrum it is necessary to carry the calculation to
	order $H_\ast^4$, and to compute the trispectrum it is necessary to
	work to order $H_\ast^6$. These terms are not enhanced by large
	infra-red effects, and consequently are not captured by resummation in
	powers of logarithms. This effect was observed in a concrete calculation
	of non-Gaussian correlation functions by Rigopoulos \& Shellard
	\cite{Rigopoulos:2005ae}.
	
	In calculating the details of the hard sub-process, the
	so-called \emph{factorization scale} $k_F$ plays a special role; it
	acts as a cut-off on the momenta of those quanta which can dress
	the final state, or circulate within loops in its interior.
	One therefore finds infra-red terms which are at most of order
	$H_\ast^2 \ln k_\ast/k_F$ or
	$\epsilon_\ast \ln k_\ast/k_F$
	and are highly suppressed by a judicious choice of $k_F$.
	The factorization scale acts as a division between the perturbative
	and non-perturbative parts of the calculation.
	In choosing it, orders of magnitude are
	important but small factors can be reshuffled between the
	various orders of perturbation theory; in general, we can think of the
	correct prescription to be a choice of the factorization scale
	to be at the Hubble scale of interest.
	
	Can the factorization scale be assigned a physical interpretation?
	It corresponds to the coarse-graining scale in the conventional
	formulation of stochastic inflation.
	In a correct calculation all dependence on $k_F$ drops out,
	but from the point of view of the subprocess within the $k_F$-sized box
	it is possible to think of it as the generation
	of a non-perturbative mass on very large scales, as originally
	pointed out by Starobinsky \& Yokoyama \cite{Starobinsky:1994bd}.
	Indeed,
	the existence of a non-perturbative mass of this type
	for the photon was suggested by Davis et al. \cite{Davis:2000zp},
	who used the Hartree approximation to study back-reaction from a
	light scalar field (see also Ref.~\cite{Sloth:2006az}).
	This result was subsequently reproduced in
	Ref.~\cite{Prokopec:2007ak}, using a more detailed analysis.
	A similar non-perturbative mass of order the Hubble scale
	was recently encountered by Riotto \& Sloth,
	who used the Fokker--Planck equation~\eref{eq:starobinsky-fokker}
	to resum diagrams containing large infra-red effects
	in an $O(N)$-symmetric scalar field theory \cite{Riotto:2008mv}.
	Riotto \& Sloth interpreted this mass in terms of screening
	of fluctuations on scales much larger than the Hubble distance.
	In the language of the
	present paper this is equivalent to choosing the factorization
	scale at the horizon.
	It is very interesting that this phenomenon is a striking parallel
	of a key aspect of hadron phenomenology, the existence of a
	saturation scale $Q_s$ of order the inverse coupling,
	$\alpha_s^{-1}$ (see, eg., Ref.~\cite{McLerran:2008uj,McLerran:2008es}),
	which defines a correlation length below which a parton observes
	a coherent background colour
	field. In both cases the interpretation is the same.

	\subsection{Fragmentation functions and superhorizon evolution}

	The outcome of the hard subprocess is not observable,
	in the hadron and de Sitter cases equally, but must be evolved to
	observable scales using the fragmentation functions
	associated with the third phase of the ``collision.''
	This evolution is described by a process analogous
	to the radiative cascade, which in the hadronic case leads to a
	DGLAP equation with the form of Eq.~\eref{eq:DGLAP} but
	with the ``spacelike'' splitting functions appropriate for initial
	state radiation replaced by ``timelike'' splitting functions
	appropriate for final state radiation. As remarked in
	\S\ref{sec:hadrons}, these two sets of splitting functions are equal
	to leading order in $\alpha_s$ but differ at higher orders.
	
	In the de Sitter case it is not presently necessary to be so
	sophisticated and the discussion can be simplified considerably.
	In particular, since we are observing at most
	$\Delta N \lesssim 10^2$ e-folds after the hard subprocess has
	taken place we can ignore processes associated with the
	perturbative scale $H_\ast^2$, which is typically tiny,
	and concentrate on those
	associated with the slow-roll scale $\epsilon_\ast \sim 10^{-2}$.
	Moreover, to avoid complications,
	let us agree that correlation functions of the fundamental fields
	are ``observable'' when evaluated at the end of inflation.
	In practice, the input to standard cosmological perturbation theory
	would be the correlation functions, evaluated at horizon re-entry,
	for the curvature perturbation and
	some number of isocurvature modes, but such issues play no role in
	the resummation of infra-red effects and can easily be accommodated
	by standard methods.

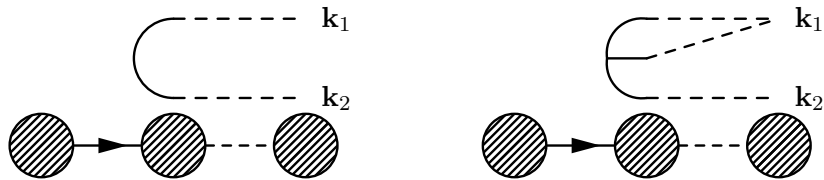
\begin{figure}
		\begin{center}
			\vspace{5mm}
			\begin{fmfgraph*}(100,60)
				\fmfleft{l}
				\fmfright{r,a,b}
				\fmf{fermion}{l,p}
				\fmf{plain,left}{q,t}
				\fmf{dashes}{q,b}
				\fmf{dashes}{t,a}
				\fmf{dashes}{p,r}
				\fmfforce{(0,0.2h)}{l}
				\fmfforce{(0.5w,0.2h)}{p}
				\fmfforce{(0.5w,0.5h)}{q}
				\fmfforce{(0.5w,h)}{t}
				\fmfforce{(w,h)}{a}
				\fmfforce{(w,0.5h)}{b}
				\fmfforce{(w,0.2h)}{r}
				\fmffreeze
				\fmfv{decoration.shape=circle,decoration.filled=shaded,
				      decoration.size=0.4h}{l,p,r}
				\fmfv{label=$\small\vect{k}_1$,label.angle=0}{a}
				\fmfv{label=$\small\vect{k}_2$,label.angle=0}{b}
			\end{fmfgraph*}
			\hspace{25mm}
			\begin{fmfgraph*}(100,60)
				\fmfleft{l}
				\fmfright{r,a,b}
				\fmf{fermion}{l,p}
				\fmf{plain,left=0.5}{q,v}
				\fmf{plain,left=0.5}{v,t}
				\fmf{plain}{v,s}
				\fmf{dashes}{p,r}
				\fmf{dashes}{q,a}
				\fmf{dashes}{t,b}
				\fmf{dashes}{s,b}
				\fmfforce{(0,0.2h)}{l}
				\fmfforce{(0.5w,0.2h)}{p}
				\fmfforce{(0.5w,0.75h)}{s}
				\fmfforce{(0.5w,h)}{t}
				\fmfforce{(0.5w,0.5h)}{q}
				\fmfforce{(w,0.2h)}{r}
				\fmfforce{(w,h)}{b}
				\fmfforce{(w,0.5h)}{a}
				\fmfforce{(w-0.5w-0.25h,0.75h)}{v}
				\fmffreeze
				\fmfv{decoration.shape=circle,decoration.filled=shaded,
				      decoration.size=0.4h}{l,p,r}
				\fmfv{label=$\small\vect{k}_1$,label.angle=0}{b}
				\fmfv{label=$\small\vect{k}_2$,label.angle=0}{a}
			\end{fmfgraph*}
		\end{center}
		\caption{\label{fig:recombine}Parton evolution and
		recombination in de Sitter,
		with time evolving from left to right in the diagram.
		Hard subprocess quanta, represented by the solid lines,
		materialize above the horizon
		and form the initial condition for the subsequent
		``third phase'' evolution, represented by the dashed lines.
		For modest evolution subsequent to horizon crossing, this
		``third phase'' evolution corresponds to a mixture of time
		dependence and recombination. If the correlation functions are
		observed following only a modest number of
		e-folds subsequent to horizon crossing, these are the only
		effects which must be taken into account.
		In general the recombination process will be very strongly
		suppressed, together with any other
		quantum processes
		associated with the scale $H_\ast^2$.
		In this approximation,
		the non-perturbative de Sitter region, represented by the
		hatched region at the bottom of the diagram, undergoes \emph{only}
		coherent time evolution from left to right.}
	\end{figure}

	With these considerations in mind
	it is possible to give a simple formula,
	exactly analogous to Eq.~\eref{eq:parton-cross-section}, for
	a correlation function of operators
	$\mathcal{O}_j$, each carrying a momentum
	$|\vect{k}_j| \gtrsim k_F$,
	within a large box of de Sitter space of size $k_0^{-1}$
	\begin{equation}
		\fl
		\langle \mathcal{O}_1(\vect{k}_1) \cdots
		\mathcal{O}_n(\vect{k}_n) \rangle_{k_0} =
		\int_{-\infty}^{\infty}
		\Big( \prod_\alpha \d \varphi^\alpha \Big) \;
		f_{k_F}(\varphi^\alpha)
		\langle \mathcal{O}_1(\vect{k}_1) \cdots
		\mathcal{O}_n(\vect{k}_n) \rangle_{k_F} 
		\Big|_{\varphi^\alpha} ,
		\label{eq:desitter-factorized}
	\end{equation}
	where the explicit $k_F$-dependence of $f(\varphi^\alpha)$ has been
	displayed, and the subscripts attached to expectation values
	denote the box size in which they are calculated.
	The correlation functions on the left- and right-hand sides of
	Eq.~\eref{eq:desitter-factorized} are to be calculated at
	a conformal time of order $\eta \sim -k_\ast^{-1}$, just after
	the time of horizon exit corresponding to the modes,
	which are assumed to be of order $k_F$.
	However, Eq.~\eref{eq:desitter-factorized} does not yet describe
	an observable.
	To do so one must take account of the two processes described in
	Fig.~\eref{fig:recombine}, corresponding to parton
	\emph{evolution} and \emph{recombination}.
	
	The most important of these processes is parton evolution,
	corresponding to resummation of $(\epsilon_\ast \nplus)^n$ terms,
	which has already been argued by many authors to correspond
	to classical time evolution
	\cite{Zaldarriaga:2003my,Lyth:2006qz,Seery:2008qj}	
	and is depicted on the left-hand side
	of Fig.~\ref{fig:recombine}. In the usual approximation, the
	``parton'' corresponding to the hatched area plus each independent
	perturbative quantum in the final state is taken to evolve
	like a separate universe. The other possible process involves
	recombination, where two or more quanta radiated during the
	hard subprocess recombine to form only a single quantum in the
	observable final state; this process is depicted on the right-hand
	side of Fig.~\ref{fig:recombine}. If $k_F \ll k_\ast$ then this process
	could receive large infra-red corrections, but will suppressed
	by powers of $H_\ast^2$ for $k_F \sim k_\ast$.
	The fragmentation functions, which describe how different final
	states resulting from the hard sub-process
	contribute to the observable of interest, can evidently be computed
	using the usual formulae of the separate universe picture of
	which the $\delta N$ formula is the most common example.
	In this picture,
	the process of recombination is described using so-called
	``$\delta N$ loops''
	whereas the simpler process of time evolution without
	recombination is described by
	the $\delta N$ tree-level terms, although both sets of
	diagrams may involve large $\epsilon_\ast \nplus$ contributions.
	
	There will be corrections to this procedure suppressed by extra
	powers of $H_\ast^2$ or $\epsilon_\ast$ compared to the
	leading-logarithm approximation, but these are usually difficult to
	compute. An example
	where it is possible to obtain one such correction explicitly
	was given by Maldacena \cite{Maldacena:2002vr},
	who studied the correlation function
	of $\langle \zeta(\vect{k}_1) \zeta(\vect{k}_2) \zeta(\vect{k}_3) \rangle$
	and gave an argument by which it could be computed in the limit
	$|\vect{k}_3| \rightarrow 0$.
	Maldacena framed his discussion in the context of a single field
	model of inflation, where $\zeta$ is conserved on superhorizon scales
	and large logarithms containing $\nplus$ are absent. In such a model,
	there is no ``phase 3'' in the collision process and we observe the
	bare outcome of the core subprocess
	(or more precisely, the effect of evolution and recombination
	is trivial), although the
	``phase 1'' radiative cascade is still operative.
	In the language of the cascade,
	a three-point function with one momentum squeezed to zero
	corresponds at leading order to
	the forbidden process of tadpole emission, and therefore vanishes.

	However, it must be remembered that the hard sub-process, here
	$\langle \zeta(\vect{k}_1) \zeta(\vect{k}_2) \rangle$, was computed in
	the wrong background. This can be accounted for by expanding in powers
	of operators carrying much softer momenta
	\cite{Bartolo:2007ti},
	which is formally the operator product
	expansion (OPE) but here is just a Taylor series. By this process
	the leading term in the emission cascade---corresponding to radiation
	of two quanta---can be isolated. The result is a correction to
	$\langle \zeta(\vect{k}_1) \zeta(\vect{k}_2) \rangle$
	of order $\epsilon H^2$, which is not enhanced by large logarithms,
	where these quantities are evaluated at
	the moment of horizon crossing corresponding to the wavenumber
	$|\vect{k}_3|$.
	This argument has been generalized
	in Refs.~\cite{Creminelli:2004yq,Cheung:2007sv}.
	In the hadron case it is known that the stochastic interpretation
	becomes less clear beyond the leading-logarithm approximation, and
	it is easier to work in terms of the OPE.
	A similar approach could presumably
	be constructed in the case of de Sitter space.
		
	\section{Discussion}
	\label{sec:discussion}
	
	In the foregoing sections it has been shown how the
	evolution of a box of de Sitter space much larger than
	the Hubble size can be given a description in the language of
	partons. In this description the vacuum expectation values of
	any light fields in the theory play the role of the Bjorken
	variable $x$, and the Hubble scale plays the role of a probe
	energy scale $Q^2$.
	In the hadron picture, $x$ characterizes the momentum fraction
	of the parton and the probe is an impinging hard particle
	which communicates a hard momentum transfer of order $Q^2$ to the
	parton. The analogue of strict Bjorken scaling in this picture
	occurs when the scalar field expectation values remain the same,
	irrespective of the scale on which the box is probed.
	Violations of strict Bjorken scaling arise from two sources:
	one is the coherent background time evolution of the box,
	but the real analogue of those processes by which Bjorken scaling
	is violated in hadrons occur when the parton radiates soft
	quanta. These can be resolved if the probe scale is sufficiently
	small. The DGLAP equation which describes the radiation process
	at leading-logarithm order
	is exactly Starobinsky's diffusion equation for the probability
	density function in stochastic inflation. It follows that
	on superhorizon scales one can
	think of time evolution in a slow-roll phase of
	de Sitter expansion as an increasing refinement of
	the scale on which features can be resolved. As we anticipated
	at the outset, this increasing refinement can be understood as a form of
	renormalization group flow.
	
	How can we reconcile the stochastic nature of the
	Starobinsky--DGLAP equation with our experience of the renormalization
	group in the ultra-violet, which ordinarily leads to screening
	of masses and coupling constants but not stochastic dynamics?
	When we use the rules of quantum field theory to compute scattering
	amplitudes or decay rates of point particles in an interacting
	theory, we imagine the bare degrees of freedom to be surrounded by
	an unresolved cloud of virtual fluctuations. An impinging particle
	which probes this cloud at some energy scale cannot distinguish
	short-lived higher-energy fluctuations within the cloud, so we must
	replace the bare degrees of freedom by the average effect
	of scattering off the unresolved cloud. When we integrate the
	renormalization group from the ultra-violet towards the infra-red
	we add new modes to the unresolved cloud and
	average over their contributions, replacing our original point
	particle description by a new one. As modes are added to the cloud,
	the masses and couplings of the point particle are slowly screened
	as increasing numbers of unresolvable quanta contribute
	incoherently to the averaging procedure.
	In the DGLAP equation, for hadrons and de Sitter equally, we are
	driving the renormalization group flow in the opposite direction.
	In these cases we begin with the existence of an unresolved cloud,
	and attempt to understand its composition as the probe moves to
	higher energy scales: in other words,
	whether we obtain screening or stochastic dynamics
	is a question of the boundary
	condition we adopt. The stochastic character of the DGLAP equation is a
	proxy for the inherent quantum mechanical uncertainty with which
	quanta become resolved from the cloud, but in either case the
	\emph{physical} picture is the same.
	
	In the renormalization group picture,
	time evolution becomes the direction of the renormalization group
	flow for the simple reason that it is time evolution which causes the
	probe scale---the Hubble scale---to vary.
	It is not ordinarily possible to understand time evolution in these
	terms, because
	evolution equations are typically second order differential equations
	and do not admit a renormalization group interpretation.
	The crucial ingredient is the slow-roll approximation, which allows
	the relevant evolution equations to be truncated to first order.
	One is immediately led to suspect that
	the underlying structure which allows such a description
	is connected to a holographic interpretation
	of de Sitter space, or the proposed ``dS/CFT correspondence.''
	There are reasons to
	believe that a dS/CFT correspondence may not exist in the same way
	as the well-established AdS/CFT correspondence, but many
	important features of the AdS/CFT holographic renormalization
	group flow are known to have analogues in de Sitter space.
	These flows have already been shown to reproduce many aspects of
	the standard theory of inflationary perturbations
	\cite{Larsen:2002et,Larsen:2003pf,vanderSchaar:2003sz,Seery:2006tq}.
	
	There appear to be significant obstructions to interpreting
	a phase of de Sitter evolution directly in terms of the parton
	evolution of some gauge field theory. Although some aspects are
	quite analogous---the occurence of a DGLAP regime, or the apparent
	existence of saturation scales associated the inverse coupling---some
	are quite different. Most strikingly, there are no degrees of freedom
	in the de Sitter picture which could play the role of Bjorken variables
	describing the gluons of an underlying gauge theory.
	The gluons are expected to play a dominant role at very high energies,
	resulting in the formation of the so-called \emph{colour glass condensate}
	\cite{Weigert:2005us,McLerran:2008uj,McLerran:2008es}.
	At these energies, multiple gluon splittings tend to fill any
	hadron wavefunction with a universal cascade of soft gluons
	carrying deeply degraded momenta. In this regime
	large effects contributed by
	$\ln x$ logarithms require the DGLAP equation to be replaced
	by the so-called Balitsky--JIMWLK equation or its relatives
	\cite{Kuraev:1977fs,Balitsky:1978ic}.
	In the case of an $O(N)$-symmetric
	scalar field theory
	in de Sitter space one could define variables more precisely
	analogous to the Bjorken $x$ by setting
	$y^\alpha = \phi^\alpha / \| \phi \|$,
	where $\| \phi \|^2 = \phi^\alpha \phi_\alpha$,
	and for which $0 \leq y^\alpha \leq 1$.%
		\footnote{This definition was suggested to me by Dmitry Podolsky.}
	A field at sufficiently
	small $y$ has a small expectation value in comparison
	with other fields whose expectation values are approaching the
	Planck scale, and in this regime one would also
	expect significant corrections
	in the de Sitter case.
	
	Although an underlying gauge theory is lacking in
	de Sitter, it is interesting that one can find a
	partial, qualitative
	analogue of the confinement phase transition by which coloured
	degrees of freedom are softly bound into colourless states
	at an energy scale around $\LambdaQCD \sim 200 \; \mbox{MeV}$.
	This is the phase transition between eternal inflation and
	slow-roll inflation towards a terminal vacuum
	\cite{Creminelli:2008es,Dubovsky:2008rf},
	which Starobinsky called the ``useful'' part of inflation in
	his original paper on the stochastic formalism
	\cite{Starobinsky:1986fx}.
	The division between these phases is marked by the
	self-reproduction scale, which is accessible to the degree that
	it sits below the Planck mass in
	many models.
	In the eternal phase, which we could perhaps loosely imagine as a
	sort of confined phase for a box of de Sitter space,
	the partons are strongly interacting and
	the trajectories of Hubble-sized regions in field space
	mix randomly. In the terminal phase
	the partons are weakly interacting
	and trajectories become
	collimated, moving on parallel paths in field space.
	There is an analogy, too, with the asymptotic freedom of QCD
	by which the theory becomes non-interacting at high energies:
	as the coupling $H_\ast^2$ decreases, interactions can be
	described increasingly well by a free scalar field theory without
	dynamical gravity \cite{Creminelli:2008es}.
	However, one should remember that all these observations are purely
	qualitative.
	
	What have we learned from the study of infra-red effects?
	There seem to be two clear conclusions, both of which have already
	been emphasized in the literature by authors working with
	different methods.
	
	Firstly, in making a prediction for what can be observed in the CMB
	one should carry out the calculation within a box that is
	in the neighbourhood of a terminal vacuum, by which is meant
	that the box as a whole (up to fluctuations described by the
	curvature perturbation) is evolving towards the hypersurface on which
	inflation ends.
	One can certainly get a different answer by
	calculating within a much larger box, but the difference arises
	because one includes in the average the possibility that some
	regions of the box are characterized by scalar vevs
	which are some way from the terminal vacuum
	\cite{Bartolo:2007ti,Enqvist:2008kt}.
	Therefore one should replace Eq.~\eref{eq:desitter-factorized}
	by a similar equation in which the distribution function $f$ is
	replaced by the probability that, given inflation is about to end,
	the scalar vevs take particular values.
	For single field inflation this probability is simply a
	$\delta$-function and the entire infra-red structure decouples from
	the theory \cite{Unruh:1998ic,Bartolo:2007ti,Enqvist:2008kt,
	Urakawa:2009my}.
	Where more than one field is present, there can be
	a non-trivial effect if inflation can end in different ways
	at different points in field space, or if one can obtain
	slightly different predictions by rolling into the same terminal
	vacuum from different directions.

	A special case of these observations applies to the calculation
	of non-Gaussian effects. Although it has been known
	for a long time that one can obtain
	significant non-Gaussianity from large scale stochastic fluctuations
	\cite{Salopek:1990re,Salopek:1990jq},
	such long wavelength phenomena
	cannot be the source of any non-Gaussian effects presently observed
	in the CMB. As new modes fall within our Hubble volume there is
	the prospect that we may eventually be able to interact with
	non-Gaussian fluctuations on very long wavelengths, but if this
	possibility exists it would seem to do so only in
	our long-term future. In making predictions for
	non-Gaussian fluctuations based on selection of specific trajectories
	in field space, one should therefore be careful that the trajectory
	does not intersect a phase of eternal inflation, and can be
	smoothly glued on to a prescription for exiting inflation.
	Likewise, if large infra-red loop corrections occur
	in making predictions for any correlation function, then
	these will generally require resummation before such correlators can be
	interpreted as observables.
	
	Secondly, since what we can observe is conditioned upon proximity
	to the end of inflation, it is clear that we cannot presently
	probe the large-scale structure generated by eternal inflation
	\cite{Lyth:2006gd,Lyth:2007jh,Bartolo:2007ti,
	Enqvist:2008kt}.
	This is like universality in the ultra-violet renormalization
	group flow, which prevents us from extrapolating the details of
	quantum gravity near the Planck mass
	from observations made at more pedestrian scales.%
		\footnote{This point has been emphasized elsewhere by
		Dmitry Podolsky; see, in particular, the discussion at
	{\scriptsize	\url{http://www.nonequilibrium.net/124-talk-munich-regularizing-correlators-curvature-perturbation/}}.}
	For example,
	if we wish to obtain top-down predictions from an \emph{a priori}
	model, perhaps obtained by construction from string theory or some
	other model of high-energy physics, then there may exist a landscape
	of vacua in which inflation can end, and the situation becomes
	very complicated. One na\"{\i}ve
	way to find a measure on this landscape
	would be to allow the scalar expectation values to diffuse over the
	landscape, and record the frequency with which they fall into
	terminal vacua. Unfortunately, even this na\"{\i}ve approach
	is a difficult undertaking in its own
	right, and could easily be complicated by other
	landscape effects
	\cite{Copeland:2007qf,Saffin:2008vi}.
	These issues were discussed concretely in
	Refs.~\cite{Podolsky:2008du,Podolsky:2008qq}.
	The role of infra-red divergences in this context
	is to reproduce the measure problem of eternal inflation.
	
	Stochastic formalisms for the purpose of computing the non-Gaussianity
	from inflation were in use throughout the 90s
	\cite{Gangui:1993tt,Gupta:2002kn}
	and more recently have been revived by
	Rigopoulos, Shellard \& van Tent
	\cite{Rigopoulos:2004gr,Rigopoulos:2005xx,Rigopoulos:2005ae}
	as a means
	to account for non-linear time evolution of the curvature perturbation
	in models where isocurvature fluctuations may be present
	(see also Refs.~\cite{Hattori:2005ac,Kunze:2006tu}).
	They therefore constitute an alternative to the popular
	$\delta N$ formula, although both are based on a
	separate universe picture \cite{Lyth:2005fi,Rigopoulos:2003ak}.
	The predictions of the stochastic formalism reproduce
	those of the $\delta N$ formula
	for time evolution,
	since they both amount to a resummation of terms of the form
	$(\epsilon_\ast N)^n$. It is now easy to see that the argument of
	\S\S\ref{sec:desitter}--\ref{sec:predictions} requires resummation of
	the $(H_\ast^2 N)^n$ terms in the $\delta N$ formula to reproduce
	the stochastic component of
	Refs.~\cite{Rigopoulos:2004gr,Rigopoulos:2005xx,Rigopoulos:2005ae}.
	We conclude that these two methods for computing non-Gaussian
	correlation functions can be regarded as completely equivalent.
	In particular, this apparently implies that the inclusion of
	a stochastic source does not lead to large non-Gaussian signals
	unless one has implicitly passed to
	a box in which infra-red effects are important, either because of
	its large size or because of an interaction with eternal inflation.
		
	\ack
	I would like to thank the participants at the workshop on
	\emph{Non-gaussianity from fundamental physics} (hosted by
	the Centre for Theoretical Cosmology, Cambridge, in September 2008)
	for many interesting discussions.
	In particular, some of the ideas discussed in this paper were
	developed during conversations with
	Peter Adshead, Neil Barnaby, Nicola Bartolo, Chris Byrnes, Xingang Chen,
	Emanuela Dimastrogiovanni, Richard Easther, Eugene Lim,
	Yeinzon Rodr\'{\i}guez, Sarah Shandera, Martin Sloth,
	Andrew Tolley,
	Cesar Valenzuela-Toledo, and Filippo Vernizzi.
	I would especially like to thank Eugene Lim for helping me locate a copy
	of Ref.~\cite{Starobinsky:1986fx}.
	
	Neil Barnaby, Sarah Shandera, Martin Sloth
	and especially Dmitry Podolsky
	offered valuable suggestions and criticisms of earlier versions of
	this paper.

	\appendix
	
	\section{Starobinsky's theory}
	\label{sec:starobinsky}

	As originally formulated,
	Starobinsky's proposal of stochastic inflation can be thought of as
	a means to account for the back-reaction
	of fluctuations in a theory of a scalar field coupled to gravity in
	de Sitter space.
	Consider the Heisenberg operator, $\Phi$,
	corresponding to any light scalar
	degree of freedom. In the stochastic prescription, this operator
	is coarse-grained according to the prescription
	\cite{Starobinsky:1986fx}
	\begin{equation}
		\fl
		\Phi = \bar{\Phi}(t,\vect{x}) +
			\int \frac{\d^3 k}{(2\pi)^3} \; \vartheta(k - \varepsilon
			a H) \left[
				a_{\vect{k}} \varphi_k(t)
				\e{-\im \vect{k} \cdot \vect{x}} +
				a_{\vect{k}}^{\ast} \varphi_k^\ast(t)
				\e{\im \vect{k} \cdot \vect{x}}
			\right] + \delta \phi ,
		\label{eq:coarse-grain}
	\end{equation}
	where $\bar{\Phi}$ is a long wavelength mean field,
	assumed to evolve according to the classical equations of motion,
	which contains fluctuations much longer than the Hubble scale
	(that is, $k \ll aH$),
	and $\delta \phi$ accounts for higher order corrections
	which are neglected. The short-wavelength integral term is treated
	according to the usual rules of quantum field theory, in such a way
	that $\{ a_{\vect{k}}, a_{\vect{k}}^\ast \}$ are annihilation and
	creation operators for modes of wavenumber $\vect{k}$ and the
	wavefunctions $\{ \varphi_k \e{-\im \vect{k}\cdot\vect{x}},
	\varphi_k^\ast \e{\im \vect{k}\cdot\vect{x}} \}$
	are solutions to the equation of motion in the interaction picture.
	The quantity $\vartheta(z)$ is Heaviside's step function, equal to
	unity for $z>0$ and zero for $z<0$, and with an indeterminate value
	at $z = 0$ whose meaning will become clear below.
	Eq.~\eref{eq:coarse-grain} is written in terms of a
	coarse-graining parameter
	$\varepsilon < 1$ (not to be confused with the slow-roll
	parameter $\epsilon \equiv -\dot{H}/H^2$)
	which can be interpreted in the DGLAP picture as
	a measure of the factorization scale, according to the rule
	$k_F = \varepsilon k_\ast$. The aim is to obtain a probability
	distribution for the long wavelength field $\bar{\Phi}$, after which
	correlation functions formed out of $\Phi$ reproduce the
	prescription given in Eq.~\eref{eq:desitter-factorized}.
	Such correlation functions do not depend on $\varepsilon$ if
	contributions from both $\bar{\Phi}$ and the short wavelength
	quantum mechanical part are kept.%
		\footnote{In applications of the stochastic formalism
		it is often the case
		that one wishes to identify large contributions from the mean
		field $\bar{\Phi}$, in which case it may be reasonable to neglect
		contributions from the hard subprocess.
		However, if this is done
		then the resulting correlation functions will contain a
		sensitivity to the coarse-graining scale and in view of the
		very large scales on which these correlation functions apply one
		should be wary of interpreting them
		as observable quantities.
		This is the counterpart of the principle, within the
		DGLAP framework, that the distribution
		function $f$ is not itself observable.

		Indeed, in
		Eq.~\eref{eq:desitter-factorized} all information about
		presently observable modes is contained within the
		hard sub-process and the infra-red resummation in
		the distribution function $f$ has the comparatively limited role of
		averaging over different vacua.
		It would seem that a similar interpretation
		should be applied to the coarse-graining scale.}
	
	An equation of motion for $\bar{\Phi}$ can be determined by substituting
	the full Heisenberg field $\Phi$ into its equation of motion,
	and using the slow-roll condition to delete the double-derivative term,
	which yields
	\begin{equation}
		3 H \dot{\bar{\Phi}} =
			- \frac{\partial V(\bar{\Phi})}{\partial \bar{\Phi}} + f ,
		\label{eq:langevin}
	\end{equation}
	where $H \equiv \dot{a}/a$ is the Hubble parameter, $V = V(\phi)$ is
	the potential.
	In Eq.~\eref{eq:langevin}
	$\alpha$ is a stochastic term, which (after accounting for the finite
	volume of space and time) is subject to the rule
	\begin{equation}
		\langle \alpha(t) \alpha(t') \rangle =
		3H^2 \left( \frac{H}{2\pi} \right)^2 \delta(t - t') .
		\label{eq:f-twopoint}
	\end{equation}
	
	The stochastic source means that the long-wavelength field does not
	evolve homogeneously, but rather develops fluctuations from place to
	place.
	From Eq.~\eref{eq:f-twopoint} one can show that the large scale
	distribution of $\phi$ is governed by a probability distribution
	$f(\bar{\Phi})$ which evolves according to a Fokker--Planck equation,
	\begin{equation}
		\frac{\partial f}{\partial t} =
		\frac{1}{3H} \frac{\partial (V' f)}{\partial \bar{\Phi}} +
		\frac{H^3}{8\pi^2} \frac{\partial^2 f}{\partial \bar{\Phi}^2} .
		\label{eq:fokker-planck}
	\end{equation}
	
	\section{Starobinsky's Fokker--Planck equation}
	\label{sec:fokkerplanck}
	In this Appendix, the derivation of the Fokker--Planck equation
	Eq.~\eref{eq:fokker-planck} is briefly recalled. The process of obtaining
	a Fokker-Planck equation from a Langevin equation
	[such as Eq.~\eref{eq:langevin}] has generated a very large literature.
	(See, for example, Zinn-Justin \cite{ZinnJustin:2002ru}.)
	Here we contrast two especially useful approaches, the first of which is
	based on the stochastic calculus of It\={o} and
	Stratonovitch, and an alternative which
	is based on a path integral.
	
	\subsection{Derivation from stochastic calculus}
	Consider the Langevin equation, Eq.~\eref{eq:langevin}, which can be
	rewritten in terms of a random variable $\theta(t)$ which is normally
	distributed,
	\begin{equation}
		\dot{\phi} = - \frac{V'}{3H} + \frac{H^{3/2}}{2\pi} \theta ,
	\end{equation}
	where $\langle \theta(t) \rangle = 0$ and
	$\langle \theta(t) \theta(t') \rangle = \delta(t-t')$. Equivalently,
	\begin{equation}
		\d \phi = - \frac{V'}{3H} \, \d t +
		\frac{H^{3/2}}{2\pi} \d \Theta ,
		\label{eq:sde}
	\end{equation}
	where $\d \Theta \equiv \theta \, \d t$ is a so-called
	\emph{Wiener process}. In this form, Eq.~\eref{eq:sde} is an example
	of a stochastic differential equation, and $\d \phi$ is sometimes
	referred to as an It\={o} process, or a generalized Wiener process.
	It is an important theorem in the
	study of stochastic processes that $(\d \Theta)^2 = \d t$ with
	probability one. This remarkable result can be justified using
	It\={o}'s theory of integration with respect to martingales,
	which allows the construction of solutions to stochastic
	differential equations such as Eq.~\eref{eq:sde} by quadrature
	(as in conventional calculus).
	However, the physical content of the statement that $(\d \Theta)^2 = \d t$
	is considerably clearer in the path integral
	context to be described in \S{B.2}
	below.
	
	We are interested in the probability density
	for $\phi(t)$ to take a value in some prescribed range
	$(\varphi,\varphi + \d \varphi)$ at time $t$.
	This probability density is labelled
	$f[\phi(t) = \varphi]$,
	The probability that $\phi(t)$ lies
	in any set $\borel$ can be written as an expectation over the indicator
	function $I[\phi(t) \in \borel]$,
	\begin{equation}
		f[\phi(t) \in \borel] \equiv
		\E\Big\{ I[\phi(t) \in \borel] \Big\} ,
		\label{eq:feynman-kac}
	\end{equation}
	which is sometimes known as the \emph{Feynman--Kac formula}.
	In the present case, this simply reads
	$f[\phi(t) = \varphi] = \E\delta[ \phi(t) - \varphi ]$.
	Now consider how the probability density varies with time. It follows
	immediately that during a small interval $\d t$ during which
	$\phi$ undergoes some shift $\d \phi$, the change induced
	in $f$ must satisfy
	\begin{equation}
		\d f[\phi(t) = \varphi] =
		\E\Big\{ \delta'[\phi(t) - \varphi] \, \d \phi +
			\frac{1}{2} \delta''[\phi(t) - \varphi] \, (\d \phi)^2 + \cdots
			\Big\} .
		\label{eq:ito-lemma-a}
	\end{equation}
	Using Eq.~\eref{eq:sde} to substitute for $\d \phi$,
	recalling that $\langle \d \Theta \rangle = 0$ and $(\d \Theta)^2 =
	\d t$ with probability one, it follows that
	\begin{equation}
		\frac{\partial f}{\partial t} =
		\int_{-\infty}^{\infty} \d \varphi \;
		f \cdot \left\{ - \delta'[\phi(t) - \varphi] \frac{V'}{3H} +
			\delta''[\phi(t) - \varphi] \frac{H^3}{8\pi^2} \right\} .
		\label{eq:ito-lemma-b}
	\end{equation}
	One now integrates by parts. Any boundary terms that are generated
	are zero, since they involve evaluation of $f(\varphi)$
	at $|\varphi| = \infty$ and the field cannot reach infinity in
	finite time.
	Once integration by parts has been performed,
	the Fokker--Planck equation~\eref{eq:fokker-planck} is obtained
	immediately, giving Starobinsky's equation
	\begin{equation}
		\frac{\partial f}{\partial t} =
		\frac{1}{3H} \frac{\partial (f V')}{\partial \varphi} +
		\frac{H^3}{8\pi^2} \frac{\partial^2 f}{\partial \varphi^2} .
	\end{equation}
	Note that we have assumed that $H$ is a constant.
	Although this approximation is reasonable, one would nevertheless
	like to remove it and instead study the theory of a scalar field
	coupled self-consistently to gravity, in which case $H$ would
	become a function of $\phi$. In this case the noise term in the
	Langevin equation is said to become ``multiplicative,'' since the noise
	depends on the field $\phi$ for which we are trying to solve.
	Such a self-consistent theory was studied by Salopek \& Bond
	\cite{Salopek:1990jq,Salopek:1990re},
	who also give references to the earlier literature.
		
	\subsection{Derivation from a path integral}
	\label{sec:path-integral-fokker-planck}
	As an alternative to the stochastic calculus, one may represent
	the operation of taking expectation values by a path integral.
	This is directly analogous to the way one may represent a quantum or
	thermal average using path integrals, and has been explored by
	many authors; a textbook treatment can be found in
	the book by Zinn--Justin \cite{ZinnJustin:2002ru}.
	
	Our point of departure for the path integral is the same Feynman--Kac
	formula, $f(\varphi) = \E \delta[\phi(t) - \varphi]$,
	given that $\phi(t_0) = \varphi_0$.
	In terms of a path integral, this reads
	\begin{equation}
		\fl
		f[\phi(t) = \varphi \mid \phi(t_0) = \varphi_0] =
		\int [\d\theta] [\d E] \; \delta[\phi(t) - \varphi]
		\delta[E(\varphi)] \; \exp \left( -\frac{1}{2} 
			\int_{t_0}^{t} \d t' \; \theta(t')^2 \right) ,
	\end{equation}
	where the field equation for $\phi(t)$ is enforced via the
	constraint $E(\varphi) = 0$,
	\begin{equation}
		E \equiv \dot{\phi} + \frac{V'}{3H} - \frac{H^{3/2}}{2\pi} \theta(t) .
	\end{equation}
	This expression can be compared to the method of
	Ref.~\cite{Podolsky:2008du}.
	After changing variable from $E$ to $\varphi$ and writing $\delta(E)$
	with the aid of an auxiliary field $\lambda$, this is the same as
	\begin{eqnarray}
		\fl\nonumber
		f(\varphi \mid \varphi_0) =
		\int [\d \theta] [\d \phi]_{\phi(t_0) = \varphi_0}^{\phi(t) = \varphi}
		[\d \lambda] \,
		\left| \det \frac{\delta E}{\delta\phi} \right|
		\\ \nonumber \mbox{} \times
		\exp\left\{ -\frac{1}{2} \int_{t_0}^{t} \d t' \;
			\left[ \theta^2 - 2 \im \lambda \left(
				\dot{\phi} + \frac{V'}{3H}
				- \frac{H^{3/2}}{2\pi} \theta \right)
			\right]
		\right\} .
	\end{eqnarray}
	The determinant $\delta E/\delta \phi$ can be written
	\begin{equation}
		\fl
		\det \frac{\delta E}{\delta \phi} =
		\exp \tr \ln \frac{\d}{\d t} \left\{
			\delta(t-t') + \vartheta(t-t') \frac{V''}{3H}
		\right\} =
		\exp \vartheta(0) \int_{t_0}^{t} \d t' \; \frac{V''}{3H} ,
	\end{equation}
	in which the determinant of $\d / \d t$ has been factored out and
	discarded, since it leads only to an infinite
	but irrelevant field-independent constant.
	The overall coefficient depends on the arbitrary value which is assigned
	to the step function at zero argument, $\vartheta(0)$,
	for which one could naturally pick any number in the range $[0,1]$.
	For the present
	we make the ``forward'' choice $\vartheta(0) = 1$, before returning
	to this question at the end the section.

	After integrating out $\theta$ and $\lambda$,
	we are left with
	\begin{equation}
		\fl
		f(\varphi \mid \varphi_0) =
		\int [\d\phi]_{\phi(t_) = \varphi_0}^{\phi(t) = \varphi} \;
		\exp \left\{
			-\frac{1}{2} \int_{t_0}^t \d t' \;
			\left[ \frac{4\pi^2}{H^3} \left( \dot{\phi} + \frac{V'}{3H}
			\right)^2 - \frac{2 V''}{3H} \right]
		\right\} .
	\end{equation}
	As in quantum mechanics this path integral has
	an equivalent representation in terms of a Schr\"{o}dinger
	equation, which can be obtained by passing to the Hamiltonian
	picture. To do so, one interprets the argument of the exponential as
	an action $S$ and defines a canonical momentum, $p$, via the rule
	$p \equiv \delta S/\delta \dot{\phi}(t)$. One obtains
	\begin{equation}
		p \equiv \frac{4 \pi^2}{H^3} \left( \dot{\phi}(t) +
			\frac{V'}{3H} \right) .
	\end{equation}
	The Hamiltonian which follows from this can be written
	\begin{equation}
		H(p,\phi) \equiv \frac{H^3}{8\pi^2} p^2 -
		\left( \frac{V'}{3H} \right) p + \frac{V''}{3H} ,
		\label{eq:hamiltonian}
	\end{equation}
	and the Schr\"{o}dinger equation is $\partial f / \partial t =
	H(-\partial/\partial\varphi,\varphi) f$, where $H$ is
	interpreted as an operator
	with all $p$s to the right of all $\varphi$s, giving
	\begin{equation}
		\frac{\partial f}{\partial t} =
		\frac{1}{3H} \frac{\partial(f V')}{\partial \varphi} +
		\frac{H^3}{8\pi^2} \frac{\partial^2 f}{\partial \varphi^2} ,
		\label{eq:fokker-planck-path-integral}
	\end{equation}
	which is again Starobinsky's equation, Eq.~\eref{eq:starobinsky-fokker}.
	
	\section{The Starobinsky--DGLAP equation}
	\label{sec:subtleties}

	In this Appendix, I briefly mention
	some subtleties in the derivation of the
	Starobinsky--DGLAP equation, Eq.~\eref{eq:desitter-DGLAP},
	which were omitted in the main text.
	
	Firstly,
	note that Eqs.~\eref{eq:desitter-DGLAP}--\eref{eq:starobinsky-fokker}
	suffer from an operator ordering ambiguity in the term
	which involves $\partial^2 V$, and
	(as discussed in~\ref{sec:fokkerplanck})
	is present no matter which method is chosen to obtain the
	Starobinsky--DGLAP equation.
	It corresponds to an arbitrary choice in discretizing
	any stochastic process, equivalent to the distinction between
	the It\={o} and Stratonovich integrals.
	In Eq.~\eref{eq:desitter-DGLAP} this
	ambiguity requires a prescription for integrating out the
	$\delta$-function, after which some fraction of the coherent background
	evolution $\delta\varphi^\alpha$ can be included in
	$f$ and the remainder in $P$. Eq.~\eref{eq:starobinsky-fokker}
	is written in the usual operator ordering convention
	originally chosen by Starobinsky, which
	corresponds to including all of this term in $P$ and none in $f$.
	This ambiguity is not expected to have significant consequences
	for inflationary predictions \cite{Matarrese:1988hc}.
	
	Secondly,
	should we expect corrections to the form of the diffusion equation
	in a more detailed treatment?
	To answer this question, it is useful
	to observe that in the continuum limit
	the Starobinsky--DGLAP equation receives \emph{no} corrections
	from higher-order connected correlation functions. Therefore the
	diffusion equation is unaffected by non-Gaussian corrections in the
	splitting functions at any order. Such corrections are present
	in loop corrections to $\Ps_\ast$, but since these
	loops are suppressed by extra powers of $H_\ast^2$ it is clear that
	non-Gaussian effects do not contribute to the leading-logarithm
	approximation.
	Another source of corrections to the DGLAP equation could have come from
	\emph{disconnected} correlation functions,
	proportional to two or more powers of a momentum-conservation
	$\delta$-function. Correlation functions of this sort could contribute
	in a leading-logarithm approximation because the extra
	$\delta$-functions each remove one power of $\delta \ln k$.
	Their effect would be to introduce higher-derivative terms into
	the Starobinsky--DGLAP equation which correct the diffusion
	term $\partial^2 f$, completely changing the character of the solutions.
	These corrections account for non-local
	transitions which are not between approximately nearest-neighbour
	Bjorken variables, and would lead to many complications.
	Such terms can also be encountered
	in diffusion-type approximations to the hadron DGLAP equation
	\cite{Cafarella:2003up}.
	
	If the tadpole condition
	$\langle \delta \phi^\alpha \rangle = 0$ is enforced then
	disconnected diagrams cannot contribute, because to reach
	leading-logarithm order such a diagram must contain at least one tadpole.
	It is fairly clear how this is to be interpreted.
	Recall that
	multiplication by a product of disconnected diagrams can be
	thought of as shifting the vacuum in which we calculate any correlation
	function. (Indeed, it is precisely to account for dressing by
	soft quanta, which arrange themselves into
	disconnected diagrams, for which we wish
	to use the DGLAP equation itself.)
	Inclusion of a background of disconnected diagrams in the
	splitting functions is tantamount to an admission that further
	vacuum redefinitions---described in the parton picture by shifts
	among the Bjorken variables---must be accounted for, which
	does not occur if the splitting functions are computed in the
	correct background.
	Accordingly, we can conclude that
	higher-derivative corrections to the DGLAP equation are absent
	in the leading-logarithm approximation.
		
	\end{fmffile}

	\section*{References}
	
	\providecommand{\href}[2]{#2}\begingroup\raggedright\endgroup

\end{document}